\documentclass[a4paper, 11pt]{article}

\usepackage[english]{babel}
\usepackage[utf8x]{inputenc}
\usepackage{ulem}
\usepackage[T1]{fontenc}
\usepackage{subfiles}

\usepackage{amsmath, amsthm, amsfonts, amssymb, bbm, algorithm, algorithmic,mathrsfs,dsfont}

\usepackage{xr-hyper} 
\usepackage[table,dvipsnames]{xcolor}
\usepackage{graphicx, caption, subcaption, natbib, enumitem, hyperref, minitoc, pgf,tikz,makecell, rotating}
\usepackage[export]{adjustbox}

\usetikzlibrary{shapes.geometric,arrows,mindmap,fit,patterns,decorations.pathreplacing}

\usepackage[top=3cm, bottom=3cm, left=3cm, right=2.5cm]{geometry}

\usepackage[linecolor=blue!60!,backgroundcolor=blue!10!,textwidth=2.5cm,textsize=scriptsize]{todonotes}

\theoremstyle{remark}
\newtheorem{remark}{Remark}

\DeclareMathOperator{\argmin}{argmin}
\DeclareMathOperator{\ind}{\mathbbm{1}}

\DeclareMathOperator{\E}{\mathbb{E}}
\DeclareMathOperator{\R}{\mathbb{R}}

\renewcommand{\P}{\mathbb{P}}
\DeclareMathOperator{\FDR}{FDR}
\DeclareMathOperator{\FDP}{FDP}
\DeclareMathOperator{\TDR}{TDR}
\DeclareMathOperator{\TDP}{TDP}

\DeclareMathOperator{\Dtr}{\cD_{\text{train}}}

\DeclareMathOperator{\Dcal}{\cD_{\text{cal}}}

\DeclareMathOperator{\Dtest}{\cD_{\text{test}}}

\newcommand{\Ztr}{Z_{\text{train}}}
\DeclareMathOperator{\Otr}{\Omega_{\text{train}}}

\DeclareMathOperator{\cDNull}{\cD^0}
\DeclareMathOperator{\cDAlt}{\cD^1}

\newcommand{\EM}{\ensuremath}

\newcommand{\cB}{\EM{\mathcal{B}}}

\newcommand{\cD}{\EM{\mathcal{D}}}

\newcommand{\cF}{\EM{\mathcal{F}}}
\newcommand{\cG}{\EM{\mathcal{G}}}
\newcommand{\cH}{\EM{\mathcal{H}}}

\newcommand{\cL}{\EM{\mathcal{L}}}

\newcommand{\cZ}{\EM{\mathcal{Z}}}

\definecolor{mygreen}{rgb}{0.82, 1.0, 0.82}
\definecolor{myred}{rgb}{ 1.0, 0.84, 0.84}

\graphicspath{ {./img/} }

\tikzset{
    side by side/.style 2 args={
        line width=2pt,
        #1,
        postaction={
            clip,postaction={draw,#2}
        }
    }
    }

\renewcommand*{\thefootnote}{\fnsymbol{footnote}}

\begin{document}
\begin{center}
{\LARGE
	{  Conformal link prediction for false discovery rate control}
}
\bigskip

Ariane Marandon

{\small
{
LPSM, Sorbonne Universit\'e, 
4, Place Jussieu, 75005, Paris, France, ariane.marandon-carlhian@sorbonne-universite.fr


}
}
\bigskip


\end{center}
\bigskip

\begin{abstract}
Most link prediction methods return estimates of the connection probability of missing edges in a graph. Such output can be used to rank the missing edges from most to least likely to be a true edge, but does not directly provide a classification into true and non-existent. 
In this work, we consider the problem of identifying a set of true edges with a control of the false discovery rate (FDR). 
We propose a novel method based on high-level ideas from the literature on conformal inference. 
The graph structure induces intricate dependence in the data, which we carefully take into account, as this makes the setup different from the usual setup in conformal inference, where data exchangeability is assumed. 
The FDR control is empirically demonstrated for both simulated and real data.

\end{abstract}

\renewcommand*{\thefootnote}{\arabic{footnote}}
\setcounter{footnote}{0}


\section{Introduction}

\subsection{Problem and aim} 

Graphs (or networks) denote data objects that consist of links (edges) between entities (nodes). Real-world examples are ubiquitous and include social networks, computer networks, food webs, molecules, etc. 
A fundamental problem in network analysis is link prediction \citep{lu2011link}, where the goal is to 
identify missing links 
in a partially observed graph. 
Biological networks such as protein-protein interaction networks \citep{PPI19} or food webs \citep{foodweb20} are typical examples of incomplete networks: because experimental discovery of interactions is costly, many interactions remain unrecorded. Link prediction can be used to identify promising pairs of nodes for subsequent experimental evaluations. Other applications include friend or product recommendation \citep{li13}, or identification of relationships between terrorists \citep{clauset2008hierarchical}.

In this work, we consider a link prediction problem, where a graph with a set of vertices $V = \{1, \dots, n \}$ and a set of edges $E$ is only partially observed: namely, we observe a sample of node pairs recorded as interacting (true edges) and a sample of pairs recorded as non-interacting (false edges). The graph can be directed or undirected and self-loops are allowed. The two observed samples of node pairs make up only a part of the set of all pairs $V \times V$, and the non-observed pairs correspond to missing information, where it is not known whether there is an edge or not. The aim is to identify the true edges among the pairs of nodes for which the interaction status has not been recorded.

There exists a variety of approaches for link prediction and they are mainly divided according to two viewpoints. In \cite{benhur05, bleakley07, li13, zhang2018link}, link prediction is treated as a classification problem.  That is, examples are constructed by associating the label 1 (or 0) with all true (or false) edges. Then, a classifier is learned by using either a data representation for each edge \citep{zhang2018link}, or kernels \citep{benhur05, bleakley07, li13}. 
Another line of research views link prediction rather as an estimation issue, namely as the problem of estimating the true matrix of the probabilities of connection between node pairs. 
In this line, \cite{chiquet20} consider maximum likelihood (ML) estimation for the Stochastic Block Model (SBM) with missing links and propose a variational approach for a variety of missing data patterns.  
\cite{gaucher21} study a low-rank model and a technique based on matrix completion tools, which is also robust to outliers. \cite{mukherjee19} give an algorithm for graphon estimation in a missing data set-up. \cite{levina23} study a special case of missing data setup called egocentric or star sampling, where observations are generated by sampling a subset of nodes randomly and then recording all of their connections, and \cite{levina17} consider the case where the recording of true/false edges can be erroneous. 
Minimax results are derived in \cite{gao2016optimal} for a least squares estimator in the SBM under a uniform and known missing data pattern and in \cite{gaucher2021maximum} for the ML estimator under a more general setting.

Concretely, the output of all of these methods are scores for all missing edges, ranking them from most likely to least likely to interact. 
Such an output is satisfying when the application constrains the number of pairs of vertices to be declared as true edges to be fixed, as e.g. in e-recommendation, where we could have to recommend the top 3-best products most likely to interest the consumer. Alternatively, other practical cases may instead require a classification of the missing edges into true and false edges together with a control of the amount of edges that are wrongly declared as true (false positives). Putting the emphasis on false positives is appropriate in many contexts.
For instance, in the reconstruction of biological networks, the edges that are classified as true are then tested experimentally in a costly process, which makes it desirable for the user to avoid false positives in the selection step. 
This is increasingly true for real-world networks that are in general very sparse. The decision of declaring a missing pair as a false edge can be viewed as a type of abstention option: based on the data, we do not have enough evidence to confidently predict it as a true edge.

How to build a reliable classification procedure? 
Using an ad hoc rule like declaring as true edges the node pairs with a connection probability above the 50\% threshold, may lead to a high number of false positives since a) probabilities may not be estimated correctly and b) even if they were, the probability of making a mistake may still be high if 
there are many node pairs with moderately elevated connection probability. 

In this work, we consider the goal of identifying a subset of the missing pairs of nodes for which we can confidently predict the presence of an edge, with a guarantee on the number of edges that are falsely predicted as true. 
Our problem can be viewed as finding the appropriate threshold (not 50\%) for the connection probabilities such that the number of false positives remains below a prescribed level. The optimal threshold depends on the problem itself. In simple settings a low threshold may be satisfactory, as for instance when most connected triplets are indeed triangles. However, on a graph with much stochasticity, the exact prediction of links is a very uncertain endeavor.

The problem is formalized in terms of controlling the false discovery rate (FDR), defined as the average proportion of errors among the pairs of vertices declared to be true edges (proportion of false discoveries). More precisely, the goal is to develop a procedure such that the FDR is below a user-specified level $\alpha$, which is an error margin that represents the acceptance level for the proportion of false edges in the selection. 
The interpretation for the user is clear: if, for instance, $\alpha$ is set to 5\% and the method returns a set of 100 node pairs, then the number of non-existent edges in this set is expected to be at most 5.

\subsection{Approach}

We propose a method that takes as input the partially observed graph and, using an off-the-shelf link prediction method, returns a set of node pairs with an FDR control at level $\alpha$. The method can be seen as a general wrapper that transforms any link prediction technique into an FDR-controlling procedure. 
Crucially, even when the quality of the link predictor is not particularly good, our method provides control of the FDR.


Our approach relies on conformal inference \citep{vovk2005algorithmic, balasubramanian2014conformal}, a statistical framework that provides generic tools to rigourously estimate uncertainty for off-the-shelf ML algorithms in various tasks. In particular, conformal prediction \citep{angelopoulos2021gentle, lei2014distribution, romano2019conformalized, romano2020classification, tibshirani2019conformal} enables to build model-free confidence intervals for the output of any ML algorithm, even "black-box", valid in finite samples without any assumptions on the data distribution besides exchangeability of the observations. Another important application of conformal inference is nonparametric hypothesis testing, via the so-called \textit{conformal} $p$-value \citep{vovk2005algorithmic, balasubramanian2014conformal, bates2023testing}.
Considering a standard testing problem $H_0 \colon X \sim P_0$ for some multivariate observation $X$ and reference distribution $P_0$, 
conformal $p$-values measure statistical significance by comparing the test statistic, or \textit{score}, to a reference set, consisting of values of the score function on observations drawn from $P_0$. 
Crucially, computing a conformal $p$-value only requires to have at hand a sample from $P_0$, rather than knowing the distribution explicitly. 
Under the exchangeability of the reference scores with the test score when the null hypothesis is true, conformal $p$-values enable to build valid tests in various contexts, such as FDR control in novelty detection \citep{bates2023testing, yang2021bonus, marandon22b, liang2022integrative}, binary classification \citep{rava2021burden,jin2022selection}, or two-sample testing \citep{hu2023two}.

We propose to use this high-level idea of comparing a score to a set of reference scores representing the null hypothesis being tested,
in order to properly threshold the link prediction probabilities for FDR control. 
In our link prediction set-up, the connection probability for a pair of nodes $(i,j)$ can be seen as a score indicating the relevance of an edge between $i$ and $j$. The afore-mentioned score comparison then turns into a comparison of 
the connection probability for a non-observed pair of nodes to connection probabilities of pairs that are known to be non-existent edges. However, the setup is markedly different from previous literature, making this transposition challenging. In particular, the graph structure makes the scores dependent on each other in an intricate way, which requires 
to build the scores with care.

\paragraph{Contributions.} The contributions of this work are summarized as follows: 

\begin{itemize}
\item We introduce a novel method to control the FDR in link prediction (Section \ref{lp:sec:meth}), which extends ideas from the conformal inference literature to graph-structured data. The proposed method acts as a wrapper that transforms any off-the-shelf link prediction (LP) technique into an FDR-controlling procedure for link prediction. It is designed to provide FDR control regardless of the difficulty of the setting and of the quality of the chosen LP technique. Moreover, the ability to use any LP technique of choice, including the state-of-the-art, makes it flexible and powerful.




\item Extensive numerical experiments \footnotemark{}  
(Section \ref{lp:sec:numexp}) assess the excellent performance of the approach and demonstrate its usefulness compared to the state of the art.
\end{itemize}
\footnotetext{
We publicly release the code of these experiments at \url{https://github.com/arianemarandon/linkpredconf}. We have also included a Jupyter notebook that illustrates the use of our procedure. 
}


\subsection{Relation to previous work}

\paragraph{Error rate control in statistical learning.}
Error rate control has notably been considered in novelty detection \citep{bates2023testing, yang2021bonus, marandon22b, liang2022integrative}, binary classification \citep{geifman17, angelopoulos2021learn, rava2021burden, jin2022selection}, clustering \citep{marandon22a} and graph inference \citep{rebafka22}. 
The setting closest to ours is that of binary classification, 
in the sense that here
the goal is to classify non-observed pairs of nodes as a 'true' or 'false' edge, given that we observe part of both true edges and non-existent edges.
In this line, some 
approaches (e.g., \citealp{zhang2018link}) view link prediction as a binary classification problem. These approaches use the graph structure to produce edge embeddings, i.e. data objects representing an edge, that are fed to a classifier as learning examples along with labels corresponding to existence or non-existence. 
The methods introduced in \cite{geifman17, angelopoulos2021learn, rava2021burden, jin2022selection} in the context of general binary classification all provide finite-sample guarantees, but the approaches and the type of guarantees vary. To be more precise, the algorithms in \cite{rava2021burden, jin2022selection} control the FDR and are very close to the conformal-based approach of \cite{bates2023testing}, whereas \cite{geifman17, angelopoulos2021learn} consider controlling the mis-classification error for a single new point and use certain bounds of the empirical risk with respect to the true risk. 

However, these approaches cannot be applied in our situation because here data examples are based on the graph structure and thus depend on each other in a complex way. In particular, we do not have i.i.d. data examples as assumed in the classical binary classification setting. 
In this regard, our method is related to the work of \cite{marandon22b} that extends the conformal novelty detection method of \cite{bates2023testing} to the case where the learner is not previously trained, but uses the test sample to output scores 
which makes the scores 
dependent. This is similar to our problem in the sense that here we aim to calibrate connection probabilities that depend on each other through the graph structure. 
 
\paragraph{Conformal inference applied to graph data.}
A few recent works \citep{huang2023uncertainty, lunde2023conformal} have considered the application of conformal prediction to graph data. However, conformal prediction is concerned with constructing prediction sets, rather than error rate control as considered here. Moreover, in these works, the prediction task concerns the nodes: \cite{huang2023uncertainty} considers node classification, while \cite{lunde2023conformal} studies prediction of node covariates (also called network-assisted regression). By contrast, in our work, the prediction task concerns the edges, and therefore, the specific dependency issue that arises differs from \cite{huang2023uncertainty, lunde2023conformal}. 
Finally, \cite{luo2021anomalous} employed conformal $p$-values to detect anomalous edges in a graph. However, their method relies on edge-exchangeability, which is a restrictive assumption, and the guarantee is only for a single edge.

\paragraph{Conformal inference for missing data problems.}   
Link prediction can be seen as a type of missing data problem where the goal is to give a prediction \textit{for} missing values, rather than \textit{with} missing values. The latter setting
was addressed within conformal prediction e.g. in \cite{zaffran2023conformal} for a regression task with missing values in the covariates and in \cite{shao2023distribution} for a particular task called matrix prediction. 
A concurrent work by \cite{gui2023conformalized} investigated conformal prediction for the matrix completion task, in which the aim is to fill in a matrix that has missing values. 
In particular, the method proposed therein is shown to provide 
coverage in a general missing data setup where the entries of the sampling matrix are either i.i.d. Bernoullis variables or known independent Bernoullis variables. \cite{gui2023conformalized} also cover the case where the sampling is unknown and non-uniform by providing error bounds for a weighted version of their method, based on tools from conformal prediction under covariate shift \citep{tibshirani2019conformal}. 
However, as outlined previously, conformal prediction is a different aim from the one considered here, which is FDR control.  

\paragraph{Link with multiple testing.}
The FDR criterion is a staple of multiple testing, 
where recent works on knockoffs and conformal $p$-values \citep{barber15, weinstein2017power, bates2023testing, yang2021bonus, marandon22b} have provided model-free procedures that come with an FDR control guarantee in finite samples. However in this work, while we do use tools of \cite{bates2023testing}, our setting does not strictly conform to a known multiple testing framework such as the $p$-value framework \citep{BH1995} (the hypotheses being random) or the empirical Bayes framework \citep{efron01, sun07} (the number of hypotheses being itself random). Hence, previous theory in that area cannot be applied.

\section{Problem setup} \label{lp:sec:setup}

\begin{figure}
\begin{minipage}{0.5\linewidth}
\centering
\begin{tikzpicture}
\begin{scope}[every node/.style={circle,thick,draw}]
\node (1) at (0,2) {1};
\node (2) at (0,0) {2};
\node (3) at (1.5,-0.75) {3};
\node (4) at (1.5, 1.) {4};
\node (5) at (-2, -0.15) {5};
\end{scope}
\draw [thick] (1) edge (2);
\draw [thick] (1) edge (4);
\draw [thick] (2) edge (3);
\draw [thick] (2) edge (5);
\draw [thick] (3) edge (5);

\draw [thick] (2) edge (4);

\end{tikzpicture}
\subcaption{Ground truth $A^*$}
\end{minipage}%
\begin{minipage}{0.5\linewidth}
\centering
\begin{tikzpicture}
\begin{scope}[every node/.style={circle,thick,draw}]
\node (1) at (0,2) {1};
\node (2) at (0,0) {2};
\node (3) at (1.5,-0.75) {3};
\node (4) at (1.5, 1.) {4};
\node (5) at (-2, -0.15) {5};
\end{scope}
\draw [thick] (1) edge (2);
\draw [thick] (1) edge (4);
\draw [thick] (2) edge (3);
\draw [thick] (2) edge (5);
\draw [thick] (3) edge (5);
\draw [dashed] (2) edge (4);
\draw [dashed] (4) edge (3);
\end{tikzpicture}
\subcaption{Observed graph $A$}
\end{minipage}

\caption{Illustration of the learning problem. The left panel shows the true complete graph $A^*$, which is not observed. The right panel describes our observation: the true edges $(1,2)$, $(1,4)$, $(2,3)$, $(2,5)$, $(3,5)$ are observed, along with the non-existent edges $(1,3)$,$(1,5)$, $(4,5)$ but the information concerning the pairs $(2,4)$ and $(3,4)$ is missing. We aim to decide for $(2,4)$ and $(3,4)$ whether there is a true edge or not.}
\label{lp:fig:setting}
\end{figure}
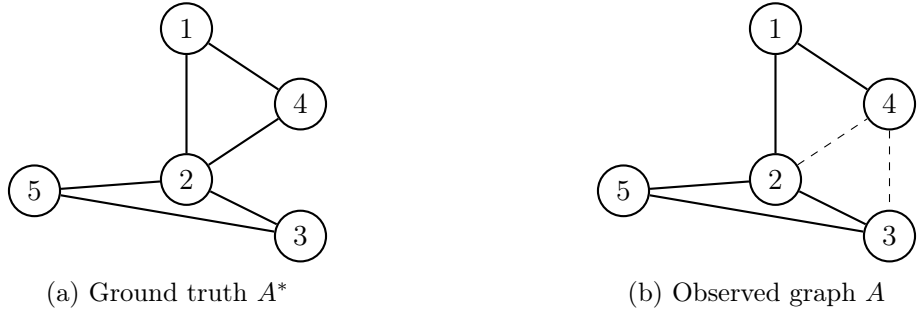

Let $A^* = (A^*_{i,j})_{1 \leq i,j \leq n}$ be the adjacency matrix of the true complete graph $\cG$, $X \in \R^{n \times d}$ a matrix of node covariates (if available), and $\Omega = (\Omega_{i,j})_{1 \leq i,j \leq n}$ the sampling matrix such that $\Omega_{i,j} = 1$ if the interaction status (true/false) of $(i,j)$ is observed, and $0$ otherwise. 
We denote by $A$ the observed adjacency matrix with $A_{i,j} = \Omega_{i,j} A^*_{i,j}$. Thus, $A_{i,j}=1$ indicates that there is an observed true edge between $i$ and $j$, whereas $A_{i,j}=0$ indicates either the observed lack of an edge or an unreported edge. The sampling matrix $\Omega$ is assumed to be observed, so that it is known which zero-entries $A_{i,j}=0$ correspond to observed false edges and which ones correspond to missing information. 
The general setting is illustrated in Figure \ref{lp:fig:setting}. 
We denote by $P$ the joint distribution of $Z^* = (A^*, X, \Omega)$, 
$Z$ the observation $(A, X, \Omega)$ and $\cZ$ the observation space. 

We next lay out our specific modelling assumptions regarding $P$. In this work, we make assumptions on the generation of the sampling matrix only. We start by reviewing the main settings encountered in the literature in this regard. 
As outlined in \cite{chiquet20}, most works on link prediction do not explicitly model the sampling matrix $\Omega$. 
To fill this gap, \cite{chiquet20} provided a formal taxonomy of missing data patterns in networks by adapting the theory developed in \cite{little2019statistical}, dividing them into three cases:
\begin{itemize}
\item Missing completely at random (MCAR): $\Omega$ is independent from $A^*$. 
\item Missing at random (MAR): $\Omega$ is independent from the value of $A^*$ on the unobserved part of the network. 
\item Missing not at random (MNAR): when the setting is neither MCAR nor MAR. 
\end{itemize}  
\cite{chiquet20} give several examples for each case. For instance, the MCAR assumption trivially includes the setting of random-dyad sampling where entries of $A^*$ are missing uniformly at random, that is considered in \cite{gao2016optimal} and is also well studied in the matrix completion literature \citep{candes2012exact, candes2010matrix, chatterjee15}. In the specific context of link prediction, another typical MCAR sampling pattern is star (or egocentric) sampling \citep{levina23}, that consists in randomly selecting a subset of the nodes and observing the corresponding row of $A^*$. A particular type of MNAR sampling is studied in \cite{gaucher21} where the sampling only depends on latent structure (e.g. node communities).  


In this work, we consider a type of MNAR sampling called double standard sampling \citep{chiquet20, sportisse2020imputation}, in which the entries of $\Omega$ are independently generated as:
\begin{align*}
\Omega_{i,j} \vert A^*, X \sim \cB( w_0 \ind_{A^*_{i,j} =0} + w_1 \ind_{A^*_{i,j} =1} ), \quad 1 \leq i, j \leq n 
\end{align*}
for some unknown sampling rates $w_0, w_1$. This amounts to consider that true edges together with false edges are missing uniformly at random at a certain status-specific rate, and can be seen as a straightforward generalization of random-dyad sampling that is more relevant for practical applications. Indeed, detecting interactions is typically more of interest rather than the detecting of non-interactions, hence one can expect that in general true edges have more chance of being reported than false edges.

\begin{figure}

\begin{minipage}{0.5\linewidth}
\centering
\begin{tikzpicture}
\begin{scope}[every node/.style={circle,thick,draw}]
\node (1) at (0,2) {1};
\node (2) at (0,0) {2};
\node (3) at (1.5,-0.75) {3};
\node (4) at (1.5, 1.) {4};
\node (5) at (-2, -0.15) {5};
\end{scope}
\draw [thick] (1) edge (2);
\draw [thick] (1) edge (4);
\draw [thick] (2) edge (3);
\draw [thick] (2) edge (5);
\draw [thick] (3) edge (5);
\draw [dashed] (2) edge (4);
\draw [dashed] (4) edge (3);
\draw [black!20](1) -- (3) node[pos=0.5, sloped] {\small$\times$}; 
\draw [black!20](4) -- (5) node[pos=0.5, sloped] {\small$\times$}; 
\draw [black!20](1) -- (5) node[pos=0.5, sloped] {\small$\times$}; 


\matrix [draw,xshift=2cm,yshift=-2cm] at (current bounding box.north east) {
  \draw [thick] (0,0) edge (0.75,0); & \node  {$\cDAlt$}; \\ 
  \draw [black!20] (0,0) -- (0.75,0) node[pos=0.5, sloped] {\small$\times$}; & \node  {$\cDNull$}; \\
  \draw [dashed] (0,0) edge (0.75,0); & \node  {$\Dtest$}; \\
};
\end{tikzpicture}
\end{minipage}
\begin{minipage}{0.5\linewidth}
\centering
\begin{tikzpicture}
\begin{scope}[every node/.style={circle,thick,draw}]
\node (1) at (0,2) {1};
\node (2) at (0,0) {2};
\node (3) at (1.5,-0.75) {3};
\node (4) at (1.5, 1.) {4};
\node (5) at (-2, -0.15) {5};
\end{scope}
\draw [thick] (1) edge (2);
\draw [thick] (1) edge (4);
\draw [thick] (2) edge (3);
\draw [thick] (2) edge (5);
\draw [thick] (3) edge (5);
\draw [black!20](1) -- (3) node[pos=0.5, sloped] {\small$\times$}; 
\draw [black!20](4) -- (5) node[pos=0.5, sloped] {\small$\times$}; 
\draw [black!20](1) -- (5) node[pos=0.5, sloped] {\small$\times$}; 

\draw [red!30!blue!70, thick] (2) edge (4);
\draw [red!30!orange!70, thick] (3) edge (4);

\matrix [draw,xshift=2cm,yshift=-2cm] at (current bounding box.north east) {
  \draw [red!30!blue!70, thick] (0,0) edge (0.75,0); & \node  {$\cH_1$}; \\
     \draw [red!30!orange!70, thick] (0,0) edge (0.75,0); & \node  {$\cH_0$}; \\ 
};
\end{tikzpicture}
\end{minipage}

\caption{Illustration of the notations introduced in Section \ref{lp:sec:setup}. The test edges $\Dtest$ (left panel) are divided into two subsets (unobserved): true edges $\cH_1$, and false edges $\cH_0$ (right panel).  } 
\label{lp:fig:notations}
\end{figure}
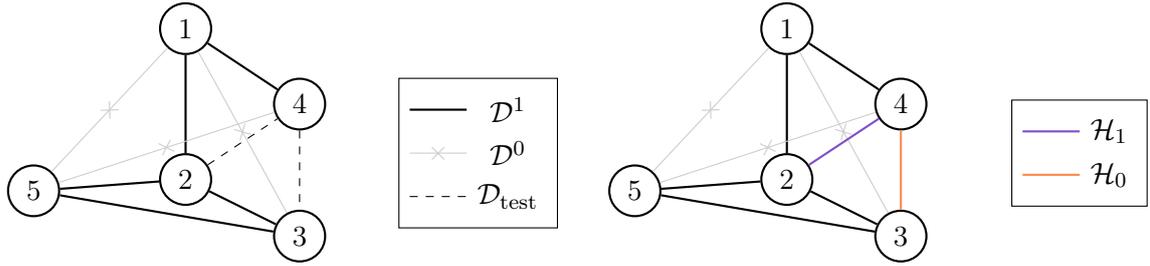

We are interested in classifying the unobserved node pairs $\{ (i,j) \; \colon \;  \Omega_{i,j} = 0\}$ into true edges and false edges, or in other words, selecting a set of unobserved node pairs to be declared as true edges, based on the observed graph structure. In order to be consistent with the notation of the conformal inference literature, we use the following notations:
\begin{itemize}
\item We denote by $\Dtest(Z)=\{ (i,j) \; \colon \;  \Omega_{i,j} = 0\}$ the set of non-sampled (or missing) node pairs and by $\cD(Z)=\{ (i,j) \; \colon \;  \Omega_{i,j} = 1\}$ the set of sampled pairs, with $\cDNull = \{ (i,j) \in \cD \; \colon \; A^*_{i,j} = 0 \}$ the set of observed non-existent edges and $\cDAlt = \{ (i,j) \in \cD \; \colon \; A^*_{i,j} = 1 \}$ the set of observed true edges. We refer to $\Dtest(Z)$ as the test set.
\item We denote by $\cH_0 = \{(i,j) \; \colon \; \Omega_{i,j} = 0, A^*_{i,j} = 0 \} $ the (unobserved) set of false edges in the test set and $\cH_1 = \{(i,j) \; \colon \; \Omega_{i,j} = 0, A^*_{i,j} = 1 \}$ the (unobserved) set of true edges in the test set. 
\end{itemize}
The notations are illustrated in Figure \ref{lp:fig:notations}. 



In our framework, a selection procedure is a (measurable) function $R=R(Z)$ that returns a subset of $\Dtest$ corresponding to the indices $(i,j)$ where an edge is declared. 
The aim is to design a procedure $R$ close to $\cH_1$, or equivalently, with $R\cap \cH_0$ (false discoveries) as small as possible. 
For any such procedure $R$, the false discovery rate (FDR) of $R$ is defined as the average of the false discovery proportion (FDP) of $R$ under the model parameter $P\in \mathcal{P}$, 
that is, 
\begin{align*}
\FDR(R)&=\E_{Z^* \sim P} [\FDP(R)],\:\:\: \FDP(R)=\frac{\sum_{i\in \cH_0} \ind_{i\in R}}{1\vee |R|}.
\end{align*}
{Similarly, the true discovery rate (TDR) is defined as the average of the true discovery proportion (TDP), that is,}
\begin{align*}
\TDR(R)&= \E_{Z^* \sim P} [\TDP(R)],\:\:\: 
\TDP(R)=\frac{\sum_{i\in \cH_1} \ind_{i\in R}}{1\vee \vert \cH_1 \vert }.
\end{align*}
Our aim is to build a procedure $R$ that controls the FDR while having a TDR (measuring the \textit{power} of the procedure) as large as possible.

\begin{remark}
In some applications, missingness can also be present in the covariate matrix $X$. 
While we do not explicitly consider this setup here,
our method can readily be applied and control the FDR over the detected edges with missingness in $X$ as long as the distributional assumption on $\Omega \vert A^*, X$ is still valid.
\end{remark}

\section{Methodology} \label{lp:sec:meth}

\subsection{Review of conformal $p$-values for out-of-distribution testing} \label{lp:sec:meth:review}

We first provide a general overview of conformal $p$-values in the classical setting of out-of-distribution testing, in which the problem is to test the null hypothesis that a data point $Z_{n+1}$ is drawn from the same (unknown) distribution $P_0$ as a i.i.d. data sample $Z_1, \dots, Z_n$. 
A conformal $p$-value \citep{heller22, bates2023testing} is a non-parametric approach that relies on reducing each multivariate observation $Z_i$ to a univariate \textit{non-conformity score} (or \textit{score} for short) $S_{i}  = g(Z_i) \in \R$, that measures the conformity to the data sample $(Z_j)_{1 \leq j \leq n}$, to evaluate the statistical evidence of being an outlier:
$$ p =  (n+1)^{-1} \left( 1+ \sum_{i=1}^n \ind \{S_i \geq S_{n+1} \} \right).$$
In words, the conformal $p$-value amounts to the rank of the test score $S_{n+1}$ among the scores of the data sample. If $(Z_j)_{1 \leq j \leq n+1}$ is i.i.d. when $Z_{n+1} \sim P_0$ and the scoring function $g$ is fixed, then $p$ is uniformly distributed under the null and hence yields a valid test. 
In practice, to use a \textit{learned} score function, one splits the data sample $(Z_j)_{1 \leq j \leq n}$ into two subsets, a \textit{training} sample $\Dtr = \{Z_1,...,Z_k\}$ and a \textit{calibration} sample $\Dcal = \{Z_{k+1}, . . . , Z_n\}$. The training sample $\Dtr$ is used to learn the score (e.g., using one-class classifiers), whereas the calibration sample $\Dcal$ is used to compute the $p$-value (hence $n$ is replaced by $n-k$ in the above equation). 
More generally, the core idea is that the validity holds as soon as we have exchangeability of the scores $S_i$ under the null. 
Moreover, in the multiple testing case, conformal $p$-values 
can be employed for FDR control when plugged into the Benjamini-Hochberg (BH) procedure \citep{BH1995}  \citep{bates2023testing, liang2022integrative, marandon22b}.

\subsection{Conformal link prediction}


Let $g: \cZ \rightarrow \R^{n \times n}$ be a \textit{scoring} function, that takes as input an observation $z \in  \cZ$, 
which is a tuple consisting of an adjacency matrix, a covariate matrix, and a sampling matrix as described in Section \ref{lp:sec:setup}, 
and returns a \textit{score} matrix $(S_{i,j})_{1 \leq i,j \leq n} \in \R^{n \times n}$, with $S_{i,j}$ estimating how likely it is that $i$ is connected to $j$.
A scoring function amounts to a link prediction algorithm (also returning in general an indicator of the relevance of an edge between $i$ and $j$ for any pair of nodes $(i,j)$) and in the sequel we use the two terms exchangeably. 
The score does not have to be in $[0,1]$: for instance, $S_{i,j}$ can be the number of common neighbors between $i$ and $j$.

\begin{algorithm}[t]
Input: {Observation $Z=(A,X,\Omega)$: (Observed) adjacency matrix $A$, node covariate matrix $X$, sampling matrix $\Omega$; LP algorithm $g$; 
sample size $\ell$ of the reference set}\\
Define $\cD=\{(i,j) : \Omega_{ij} = 1 \}$; $\Dtest=\{(i,j) : \Omega_{ij} = 0 \}$; $\cDNull = \{ (i,j) \in \cD: A_{ij} =0\}$\\
1. Sample $\Dcal$ of size $\ell$ uniformly without replacement from $\cDNull$; \\
2. Learn $g$ on ``masked'' observation $\Ztr=(A,X,\Otr)$, with $\Otr$ given by  
\begin{align*}
(\Otr)_{i,j}= \begin{cases}
      0 & \text{if}\ (i,j) \in \Dcal,  \\
      \Omega_{i,j} & \text{otherwise.}
    \end{cases}
\end{align*}
3. For each $(i,j) \in \Dcal \cup \Dtest$, compute the score $S_{i,j} = g(\Ztr)_{i,j}$ \\
4. For each $(i,j) \in \Dtest$, compute the conformal $p$-value $p_{(i,j)}$ as given in equation \eqref{lp:eq:confpvalues} \\
5. Apply BH using $(p_{(i,j)})_{(i,j) \in \Dtest}$ as input $p$-values, 
providing a rejection set $R(Z) \subset \Dtest$. \\
Output: { $R(Z)$} 
  \caption{Conformal link prediction}\label{lp:algo:main}
\end{algorithm}

To obtain a set of edges with FDR below $\alpha$, we borrow from the literature on 
conformal $p$-values \citep{bates2023testing, marandon22a} to formulate the following idea: some of the observed false edges can be used as a reference set, 
by comparing the score for a node pair in the test set to scores computed on false edges to determine if it is likely to be a false positive. 
%
Effectively, we will declare as edges the pairs that have a test score higher than a cut-off $\hat t$ computed from the calibration set and depending on the level $\alpha$. In detail, the steps are as follows:
\begin{enumerate}
\item Build a reference set $\Dcal$ of false edges by sampling uniformly without replacement from the set of observed false edges $\cDNull$


\item Run an off-the-shelf LP algorithm $g$ on the ''masked'' observation $\Ztr=(A,X,\Otr)$, where $(\Otr)_{i,j}= 0$ if $(i,j) \in \Dcal$ and $\Omega_{i,j}$ otherwise.

\item Compute the scores for the reference set and for the test set, $S_{i,j} = g(\Ztr)_{i,j}$ for $(i,j) \in \Dcal \cup \Dtest$; 
\item Compute the \textit{conformal} $p$-values $(p_{(i,j)})_{(i,j) \in \Dtest}$ given by
\begin{align} \label{lp:eq:confpvalues}
    p_{(i,j)} = \frac{1}{\vert \Dcal \vert +1} \left(1 + \sum_{(u,v) \in \Dcal} \ind \{S_{(i,j)} \leq S_{(u,v)} \} \right), 
    \quad (i,j) \in \Dtest ;
\end{align}
\item Declare as true edges the node pairs in the test set that are in the rejection set returned by the BH procedure applied to the $p$-values $(p_{(i,j)})_{(i,j) \in \Dtest}$. 
\end{enumerate}
The procedure is summarized in Algorithm \ref{lp:algo:main}. 
We now review each step one by one in order to provide the intuition behind the proposed procedure.
Step 1 builds a reference set of false edges that are sampled in such a way as to imitate the missingness of false edges in the test set, i.e. the sampling of $\cH_0$. 
Step 2 runs an off-the-shelf LP algorithm $g$ on $\Ztr$ (instead of $Z$) so as to treat the edge examples $(i,j)$ in the reference set $\Dcal$ as unreported information when learning the score, which is necessary to avoid biasing the comparison of the test scores to the reference scores.  Otherwise, the scoring $g$ may use the knowledge that the node pairs in the reference set are false edges and produce an overfitted score for those. Together with Step 1, this is crucial to fabricate good reference scores that are representative of the scores of false edges in the test set. Finally, the remaining steps correspond exactly to the conformal procedure of \cite{bates2023testing} for out-of-distribution testing and we refer the reader to Section \ref{lp:sec:meth:review} and to \cite{bates2023testing} for more details.

\begin{remark} \label{lp:rem:nullprop}
This type of procedure is designed to control the FDR at level $\frac{ \vert \cH_0 \vert} {\vert \Dtest \vert} \alpha < \alpha$. To maximize power, we recommend to apply the procedure at level $\alpha / \hat \pi_0$ where $\hat \pi_0 \in ]0,1[$ is an estimate of $\frac{ \vert \cH_0 \vert} {\vert \Dtest \vert}$. 
In particular, tools from the multiple testing literature on the estimation of the proportion of null hypotheses may be employed, e.g. by using Storey's estimator \citep{marandon22b}. 
\end{remark}

\begin{remark} Many LP algorithms (e.g. \citealp{zhang2018link}) are not trained on the entire set of observed edges $\cD$ but on a subset $\Dtr \subset \cD$ with a 50-50\% distribution of true and false edges. As most real-world networks are sparse, typically all observed edges $\cDAlt$ are used for training and a randomly chosen subset of false edges in $\cDNull$ of the same size as $\cDAlt$. Then the reference set $\Dcal$ is naturally chosen among the false edges in $\cDNull$ that are not used in $\Dtr$ for learning the predictor. Consequently, in practice choosing a reference set $\Dcal$ does not diminish the amount of data on which the predictor is learned. 
\end{remark}

\begin{remark} \label{rem:samplesize}
The sample size of the reference set $\Dcal$ must be large enough to ensure a good power, as pointed out in previous work using conformal $p$-values in the novelty detection context \citep{mary2022semi, marandon22b, yang2021bonus}. In particular, \cite{mary2022semi} give a power result under the condition that $\vert \Dcal \vert \gtrsim \vert \Dtest \vert / (k\alpha)$, where $k$ is the number of "detectable" novelties. Consequently, our recommendation is to choose $\vert \Dcal \vert$ of the order of $\vert \Dtest \vert / \alpha$; this choice works reasonably well in our numerical experiments.
\end{remark}

\subsection{Link with conformal out-of-distribution testing} 
In this section, we explicit the output of typical link prediction methods, to write each score $S_{i,j}$ as a (learned) function of an \textit{edge embedding} representing the information that is observed about the pair $(i,j)$. 
This allows to reframe our approach as using the procedure of \cite{bates2023testing} (see Section \ref{lp:sec:meth:review}) on data objects that have a non-exchangeable dependence structure.  

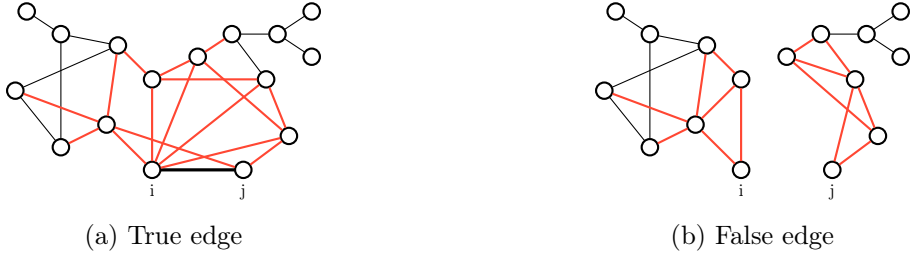
\begin{figure}

\begin{minipage}{0.5\linewidth}
\centering
\begin{tikzpicture}[scale=0.6,every node/.style={scale=0.6}]

\tikzstyle{hop} = [thick,red!80!orange!80]

\node [circle,thick,draw] (i) at (0,0) {};
\node [circle,thick,draw] (j) at (2,0) {};
\node[below] at (i.south){i};
\node[below] at (j.south){j};

\begin{scope}[every node/.style={circle,thick,draw,scale=0.6}]

\node (1) at (-1,1) {};
\node (2) at (-0.75,2.75) {};
\node (3) at (0,2) {};
\node (4) at (1,2.5) {};
\node (5) at (3,0.75) {};
\node (6) at (2.5,2) {};

\node (7) at (-2,0.5) {};
\node (8) at (-2,3) {};
\node (9) at (-2.75,3.5) {};
\node (10) at (-3,1.75) {};
\node (12) at (1.75,3) {};
\node (13) at (2.75,3) {};
\node (14) at (3.5,3.5) {};
\node (15) at (3.5,2.5) {};

\end{scope}

\draw [very thick] (i) edge (j);

\draw [hop] (1) edge (i);
\draw [hop] (1) edge (j);
\draw [hop]  (3) edge (i);
\draw [hop] (4) edge (i);
\draw [hop] (5) edge (i);
\draw [hop] (5) edge (j);
\draw [hop] (6) edge (i);

\draw [hop] (1) edge (2);
\draw [hop] (2) edge (3);
\draw [hop] (3) edge (4);
\draw [hop] (4) edge (5);
\draw [hop] (5) edge (6);
\draw [hop] (3) edge (6);
\draw [hop] (7) edge (1);
\draw (8) edge (2);
\draw (9) edge (8);
\draw (10) edge (7);
\draw [hop] (4) edge (12);
\draw (6) edge (12);
\draw (12) edge (13);
\draw (10) edge (2);
\draw (8) edge (7);
\draw (13) edge (14);
\draw (13) edge (15);
\draw [hop]  (1) edge (10);

\end{tikzpicture}
\subcaption{True edge} \label{lp:fig:embeddingtrue}
\end{minipage}%
\begin{minipage}{0.5\linewidth}
\centering
\begin{tikzpicture}[scale=0.6, every node/.style={scale=0.6}]

\tikzstyle{hop} = [thick,red!80!orange!80]

\node [circle,thick,draw] (i) at (0,0) {};
\node [circle,thick,draw] (j) at (2,0) {};
\node[below] at (i.south){i};
\node[below] at (j.south){j};

\begin{scope}[every node/.style={circle,thick,draw,scale=0.6}]

\node (1) at (-1,1) {};
\node (2) at (-0.75,2.75) {};
\node (3) at (0,2) {};
\node (4) at (1,2.5) {};
\node (5) at (3,0.75) {};
\node (6) at (2.5,2) {};
\node (7) at (-2,0.5) {};
\node (8) at (-2,3) {};
\node (9) at (-2.75,3.5) {};
\node (10) at (-3,1.75) {};
\node (12) at (1.75,3) {};
\node (13) at (2.75,3) {};
\node (14) at (3.5,3.5) {};
\node (15) at (3.5,2.5) {};

\end{scope}

\draw [hop] (1) edge (i);
\draw [hop] (1) edge (2);
\draw [hop] (1) edge (3);
\draw [hop] (2) edge (3);
\draw [hop] (3) edge (i);

\draw [hop] (4) edge (5);
\draw [hop] (4) edge (6);
\draw [hop] (5) edge (j);
\draw [hop] (5) edge (6);
\draw [hop] (6) edge (j);

\draw [hop] (7) edge (1);
\draw (8) edge (2);
\draw (9) edge (8);
\draw (10) edge (7);
\draw [hop] (4) edge (12);
\draw [hop] (6) edge (12);
\draw (12) edge (13);
\draw (10) edge (2);
\draw (8) edge (7);
\draw (13) edge (14);
\draw (13) edge (15);
\draw [hop]  (1) edge (10);

\end{tikzpicture}
\subcaption{False edge} \label{lp:fig:embeddingfalse}
\end{minipage}

\caption{Example of K-hop (K=2) neighborhood of $(i,j)$ (in color), for when (a) $i$ and $j$ are connected and (b) $i$ and $j$ are not connected. }
\label{lp:fig:embedding}
\end{figure}

For simplicity of presentation, in this section we consider an undirected graph without node covariates. 
Link prediction methods output a matrix of predictions/scores $(g (Z)_{i,j})_{1 \leq i,j \leq n}$, where the prediction for $i$ and $j$ can be written in general as 
\begin{align} 
g (Z)_{i,j} &= h( W_{i,j} ), \label{lp:eq:scorelp} \\
W_{i,j} :&= (A_{i, \bullet}, A_{j, \bullet}, A^2_{i, \bullet}, A^2_{j, \bullet}, \dots, A^K_{i, \bullet}, A^K_{j, \bullet}) \label{lp:eq:embed}, 
\end{align}
for a real-valued measurable function $h$ and a given fixed $K \in \{1, \dots, n \}$ and where $A^k_{i, \bullet}$ is defined as the $i$-th row of  $A^k$. In the \eqref{lp:eq:embed}, the r.v. $W_{i,j}$ can be thought of as the "$K$-hop neighborhood" of $(i,j)$. It represents an embedding for the node pair $(i,j)$, that describes a \textit{pattern} of connection around $i$ and $j$. If the graph has some structure, it should be observed that the pattern differs when $i$ and $j$ are connected compared to when they are not. Moreover, there should be some similarity between the patterns observed for true edges, as compared to false edges. Figure \ref{lp:fig:embedding} gives an illustration in the case of a graph with community structure. When there is an edge between $i$ and $j$ (Figure \ref{lp:fig:embeddingtrue}), $i$ and $j$ are involved in a same group of nodes that is densely connected (community). Conversely, when there is no edge between them (Figure \ref{lp:fig:embeddingfalse}), $i$ and $j$ belong to separate groups of densely connected nodes that share few links between them. 

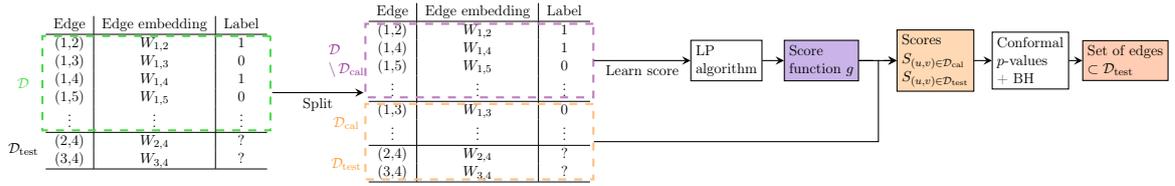
\begin{figure}
\centering
\begin{tikzpicture}[node distance = 3.5cm, scale=0.5, every node/.style={transform shape}]

\tikzstyle{box} = [rectangle, minimum width=6cm, minimum height=2.5cm, dashed]
\tikzstyle{box2} = [rectangle, minimum width=1cm, minimum height=1cm, draw=black]
\tikzstyle{arrow} = [->,>=stealth]

\node (data) at (0,0){
\begin{tabular}{c|c|c}
Edge & Edge embedding & Label \\ 
\hline
(1,2) & $W_{1,2}$ & 1 \\
(1,3) & $W_{1,3}$ & 0 \\
(1,4) & $W_{1,4}$ & 1 \\
(1,5) & $W_{1,5}$ & 0 \\
$\vdots$ & $\vdots$ & $\vdots$ \\
\hline
(2,4) & $W_{2, 4}$ & ? \\
(3,4) & $W_{3, 4}$ & ? \\
\hline
\end{tabular}
}; 
\node (cDain) [left of=data, yshift=2.5mm, text=green!80!black!70] {$\cD$};
\node (dtest)  [left of=data, yshift=-1.5cm]{$\Dtest$};
\node (cDainbox) [box, right of=cDain, draw=green!80!black!70, thick] {}; 
\node (dtestbox) [box, right of=dtest, minimum height=0.80cm] {}; 

\node (newdata)[right of=data, xshift=5cm]{
\begin{tabular}{c|c|c}
Edge & Edge embedding & Label \\ 
\hline
(1,2) & $W_{1,2}$ & 1 \\
(1,4) & $W_{1,4}$ & 1 \\
(1,5) & $W_{1,5}$ & 0 \\
$\vdots$ & $\vdots$ & $\vdots$ \\
\hline
(1,3) & $W_{1,3}$ & 0 \\
$\vdots$ & $\vdots$ & $\vdots$ \\
\hline
(2,4) & $W_{2, 4}$ & ? \\
(3,4) & $W_{3, 4}$ & ? \\
\hline
\end{tabular}
}; 
\node (cDain2) [left of=newdata, yshift=0.85cm, align=left, text=red!50!blue!80] {$\cD$ \\ $\backslash \Dcal$};
\node (dtest2)  [left of=newdata, yshift=-1.9cm, text=orange!80]{$\Dtest$};
\node (dcal)  [left of=newdata, yshift=-0.8cm, text=orange!80]{$\Dcal$};
\node (cDainbox2) [box, right of=cDain2, minimum height=1.95cm, draw=red!50!blue!50, thick] {}; 
\node (dtestbox2) [box, right of=dtest2, minimum height=0.80cm] {}; 
\node (dcalbox) [box, right of=dcal, minimum height=1cm] {}; 
\node (dtestcalbox) [box, right of=dcal, minimum height=2cm, draw=orange!50, thick, yshift=-0.5cm] {}; 

\node (algo) [box2, right of=cDainbox2, xshift=3cm, align=left] {LP \\ algorithm}; 
\node (score) [box2, right of=algo, align=left, xshift=-1cm, fill=red!30!blue!30] { Score \\ function $g$}; 
\node(mid)[right of=score, xshift=-2cm] {};
\node (compute) [box2, right of=mid, align=left, fill=orange!30, xshift=-2cm] {Scores \\ $S_{(u,v) \in \Dcal}$ \\$ S_{(u,v) \in \Dtest}$}; 
\node (CK) [box2, right of=compute, xshift=-1cm, align=left] {Conformal \\ $p$-values \\ + BH }; 
\node (out) [box2, right of=CK, align=left, xshift=-1cm, fill=red!30!orange!30] {Set of edges \\ $\subset \Dtest$};

\draw [arrow] (data)--(newdata) node[midway, below]{Split}; 
\draw [arrow] (cDainbox2)--(algo) node[midway, below]{Learn score}; 
\draw [arrow] (algo)--(score); 
\draw  (score)--(mid); 
\draw  [arrow] (mid)--(compute); 
\draw [arrow] (score)--(compute); 
\draw (dtestcalbox)-|( [yshift=0.15cm] mid.south); 
\draw [arrow] (compute)--(CK); 
\draw [arrow] (CK)--(out); 

    
\end{tikzpicture}
\caption{Sketch of the procedure proposed in this work.} \label{lp:fig:proc}
\end{figure}

For instance, in the case of the common neighbors (CN) heuristic \citep{lu2011link}, we have that $g(Z)_{i,j}= A_{i, \bullet}^T A_{j, \bullet}$. 
Alternatively, when considering supervised approaches such as binary classification-based \citep{zhang2018link, bleakley07}, maximum likelihood-based \citep{kipf2016variational, chiquet20} or matrix completion-based \citep{levina23, gaucher21}, the link prediction function can be written as the minimizer of an empirical risk (ERM): 
\begin{align} \label{lp:eq:scoreERM}
g(Z)_{i,j} = \hat h (W_{i,j}),  \text{ with } \hat h \in \left\{  \underset{h \in \cF}{\argmin} \sum_{(i,j) \in \cD 
} \cL[ h(W_{i,j}), 
A_{i,j} ] \right\}, 
\end{align}
with $\cL : [0,1] \times \{0,1\} \rightarrow \R$ a loss function and $\cF$ a function class. 
In \eqref{lp:eq:scoreERM}, $h(W_{i,j}) $ is an estimate of the probability that there is an edge between $i$ and $j$, and the error term $\cL[ h(W_{i,j}), A_{i,j} ]$ quantifies the difference between the prediction $h(W_{i,j})$ and the true $A_{i,j} $. 
The ERM formulation for the afore-mentioned supervised approaches can be justified as follows: 
	\begin{itemize}
		\item Binary classification approaches \citep{zhang2018link, bleakley07}: In that case the ERM formulation \eqref{lp:eq:scoreERM} is obvious. For instance, for SEAL \citep{zhang2018link}, $g(Z)_{i,j}$  is given by a GNN that takes as input the $K$-hop subgraph around $(i,j)$, excluding the edge between $(i,j)$ if there is one observed, and augmented with node features that describe the distance of each node in the subgraph to $i$ and to $j$. The parameters of the GNN are fitted by minimizing the cross-entropy loss over a set $\Dtr \subset \cD$ of observed true/false edges. In practice, $\Dtr$ is subsampled from $\cD$ in order to have a $50\%-50\%$ partitioning between true and false edges. 
				
	\item Maximum likelihood approaches \citep{kipf2016variational, chiquet20}: Maximum likelihood approaches aim to optimize a lower bound on the likelihood (ELBO). This lower bound is an expectation and therefore, using Monte-Carlo approximation, we end up with a function of the form \eqref{lp:eq:scoreERM}. For instance, for VGAE \citep{kipf2016variational}, $g(Z)_{i,j}$ is given by the scalar product $H_i^T H_j$ where $H_u$ is a \textit{node embedding} for node $u$, the embedding matrix $H \in \R^{n \times n}$ being the output of a GNN. It follows that $H_u = \phi(A_{u, \bullet},  A^2_{u, \bullet}, \dots, A^L_{u, \bullet})$ for some function $\phi$, with $L$ the number of layers of the GNN. 

	\item Matrix completion \citep{levina23, gaucher21}: e.g., for \cite{levina23}, one can rewrite the estimated probability matrix $\hat P$ as $\hat P =  \argmin_{P} \{ \sum_{(u,v) \in \cD} (A_{u,v} - P_{u,v})^2, P = A_{in}^T\Theta A_{in}, \; \text{rank}(\Theta) \leq r\}$ where $A_{in}$ is the sub-matrix of $A$ consisting only of the observed entries. Hence, in that case, $g(Z)_{i,j}$ is of the form \eqref{lp:eq:scoreERM} with $h(W_{i,j})=\phi(A_{i, \bullet}, A_{j, \bullet})$ for some function $\phi$. 
	\end{itemize}
In this view, our procedure proceeds by learning a binary classification rule for a set of examples indexed by $\Dtr = \cD \backslash \Dcal$, each learning example consisting of an edge embedding and its "label" (true/false), and then using a reference set of false edges examples indexed by $\Dcal$ to properly calibrate the edge probabilities for $\Dtest$. As such, it boils down to the approach proposed by \cite{bates2023testing} for FDR control except for a supervised context and with non-exchangeable data objects. Sampling $\Dcal$ in a manner that mimics the missingness for false edges and learning on the edge examples indexed by $\cD \backslash \Dcal$ allows to control the FDR despite this dependence. 
Figure \ref{lp:fig:proc} provides a high-level sketch.

\section{Numerical experiments} \label{lp:sec:numexp}

In this section, we study the performance of our method (Algorithm \ref{lp:algo:main}) both on simulated data (Section \ref{lp:sec:simu}) and on a real dataset (Section \ref{lp:sec:realdata}). 
We consider two choices for the scoring function $g$: the GNN-based method of \citep{zhang2018link} called SEAL and the Common Neighbors (CN) heuristic, yielding the procedures \texttt{CN-conf} and \texttt{SEAL-conf}. SEAL is used with a hop number of $2$, for the GNN we use GIN \citep{xu2018powerful} with 3 layers and 32 neurons, and we train for 10 epochs with a learning rate of $0,001$. 
In addition, following Remark \ref{lp:rem:nullprop}, we consider a version of our method where Algorithm \ref{lp:algo:main} is applied at level $ \alpha / \hat \pi_0$, with $\hat \pi_0$ a suitable estimator of $\vert \cH_0 \vert / \vert \Dtest \vert $. Specifically, following \cite{marandon22b}, we use Storey's estimator \citep{STS2004} given by $\hat \pi_0 =(1-\lambda)^{-1}(1+ \sum_{(i,j) \in \Dtest} \ind \{p_{i,j} \geq \lambda \} )$ with $\lambda =1/2$, which gives the procedures \texttt{CN-conf-storey} and \texttt{SEAL-conf-storey}. 
Finally, we compare the performance of our methods to a "naive" procedure for FDR control (called fixed threshold procedure hereafter) in which we select in $R(Z)$ the edges $(i,j) \in \Dtest$ for which $g(Z)_{i,j} \geq 1-\alpha$ (here we assume that $g \in [0,1]$, otherwise, scores are normalized into $[0,1]$ by standardizing the values and applying the sigmoid function). Combined with either SEAL or CN, this yields the procedures \texttt{CN-fixed} and \texttt{SEAL-fixed}.  If the probabilities $g(Z)_{i,j}$ are poorly estimated, these fixed threshold procedures are expected to not control the FDR at level $\alpha$ in general.

\subsection{Simulated data} \label{lp:sec:simu}

In this section, we evaluate our method by generating the true graph $A^*$ from two popular random graph models: the Stochastic Block Model and the graphon model \citep{matias2014modeling}. 

\subsubsection{Stochastic Block Model} \label{lp:sec:simu:sbm}
\usetikzlibrary{shapes.geometric, arrows}

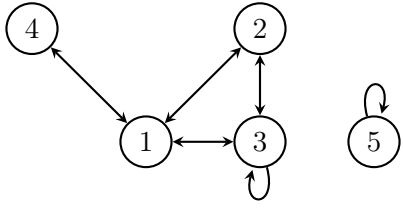
\begin{figure}
\tikzstyle{darrow} = [thick,<->,>=stealth]

\centering
\begin{tikzpicture}
\begin{scope}[every node/.style={circle,thick,draw}]
\node (1) at (0,0) {1};
\node (2) at (1.5,1.5) {2};
\node (3) at (1.5,0.) {3};
\node (4) at (-1.5, 1.5) {4};
\node (5) at (3.,0.) {5};
\end{scope}
\draw [darrow] (1) edge (2);
\draw [darrow] (1) edge (3);
\draw [darrow] (1) edge (4);
\draw [darrow] (2) edge (3);
\draw [darrow] (3) edge [loop below] (3);
\draw [darrow] (5) edge [loop above] (5);
\end{tikzpicture}
\caption{Illustration of the SBM model considered in Section \ref{lp:sec:simu}. Nodes represent classes, edges indicate connection patterns between classes. }
\label{lp:fig:sbm}
\end{figure}

To start with, $A^*$ is generated from a Stochastic Block Model:
\begin{align*}
C_1, \dots, C_n &\underset{i.i.d.}{\sim} \sum_{q=1}^Q \pi_q \delta_q \\
A^*_{i,j} \vert C_1, \dots, C_n &\sim \cB \left( \gamma_{C_i, C_j} \right), \quad 1 \leq i < j \leq n 
\end{align*}
with $Q=5$ classes, mixture proportions $\pi = (1/5, \dots 1/5)$, and connectivity matrix $\gamma$ given by
\begin{align*}
\gamma = 
\begin{pmatrix}
\epsilon &p &p &p & \epsilon \\
p & \epsilon & p & \epsilon & \epsilon \\
p &p &p &\epsilon & \epsilon \\
p &\epsilon &\epsilon &\epsilon &\epsilon \\
\epsilon &\epsilon &\epsilon &\epsilon &p 
\end{pmatrix}.
\end{align*}
In this model, the graph $A^*$ displays a variety of connection patterns including community structure and hubs, see Figure \ref{lp:fig:sbm} for an illustration. 

The overall difficulty of the problem resides within two distinct statistical tasks: learning the edge link probabilities (i.e. the score), and the multiple testing issue of controlling the false discovery rate with a given test statistic. 
The main quantities that govern the difficulty of the learning problem are the density of the graph $\rho = n^{-2}\E(\sum_{i,j} A^*_{i,j}) = \sum_{q,l} \gamma_{q,l}$ and the number of nodes $n$ \citep{gaucher2021maximum}. For the FDR control, the difficulty mainly depends on the signal-to-noise ratio $\text{SNR} = p / \epsilon$ as well as the proportion of true edges to false edges within the test set $\vert \cH_1 \vert  /  \vert \cH_0 \vert = \frac{(1-w_1) \rho}{(1-w_0)(1-\rho)}$ and the 
number of observed false edges $w_0(1-\rho)$. 
Indeed, the higher the SNR, the more true edges display different connection patterns as compared to false edges, whereas increasing the proportion of true edges to false edges within the test set and the calibration sample size improves the estimation of the false discovery rate in Algorithm \ref{lp:algo:main}, see Remark \ref{rem:samplesize}.

\begin{figure}
\centering
 \begin{tabular}{ccc}
  \multicolumn{3}{l}{ (1) $p=0.5, p/\epsilon=10$ } \\
 \includegraphics[width=0.15\linewidth]{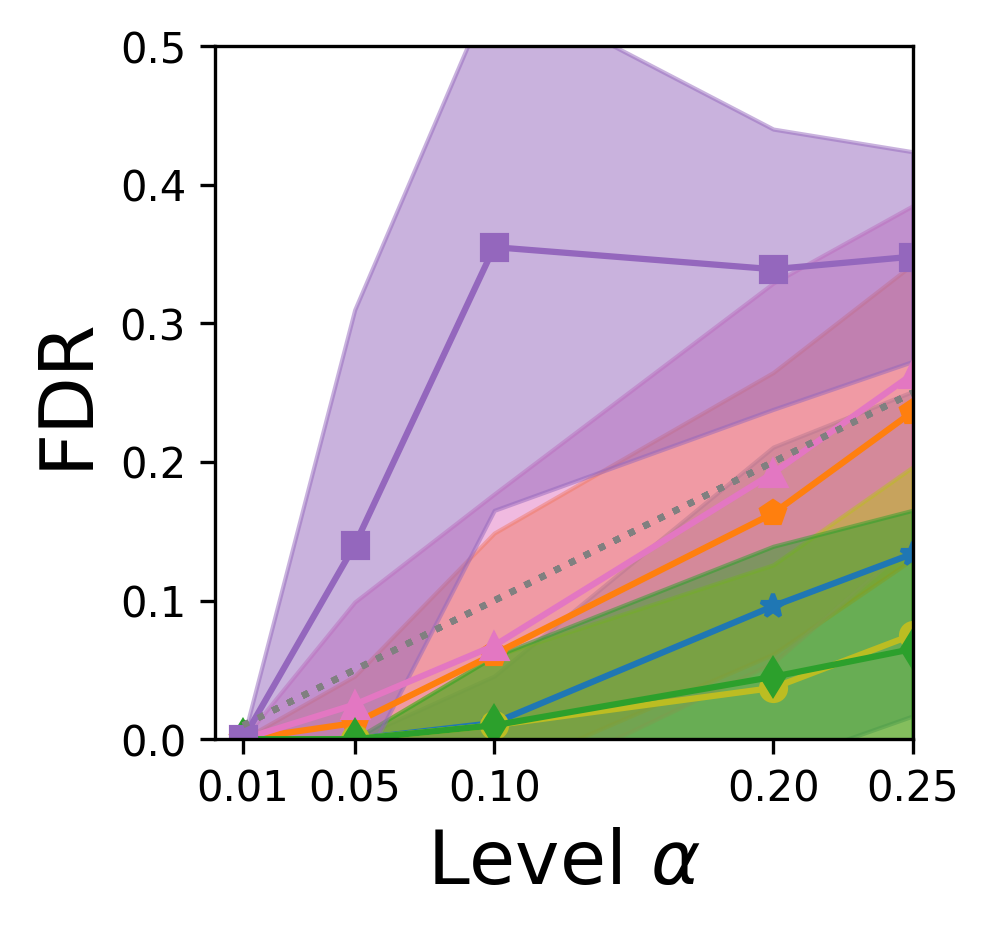}
\includegraphics[width=0.15\linewidth]{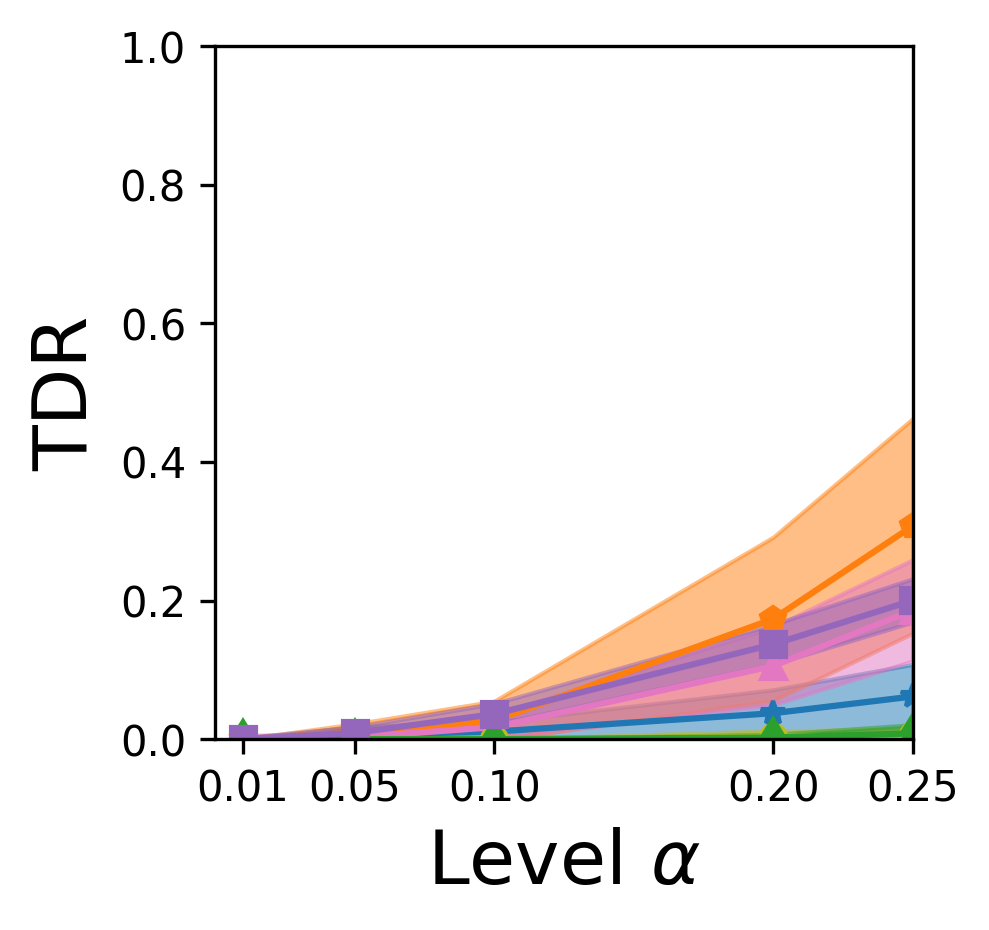}
&
\includegraphics[width=0.15\linewidth]{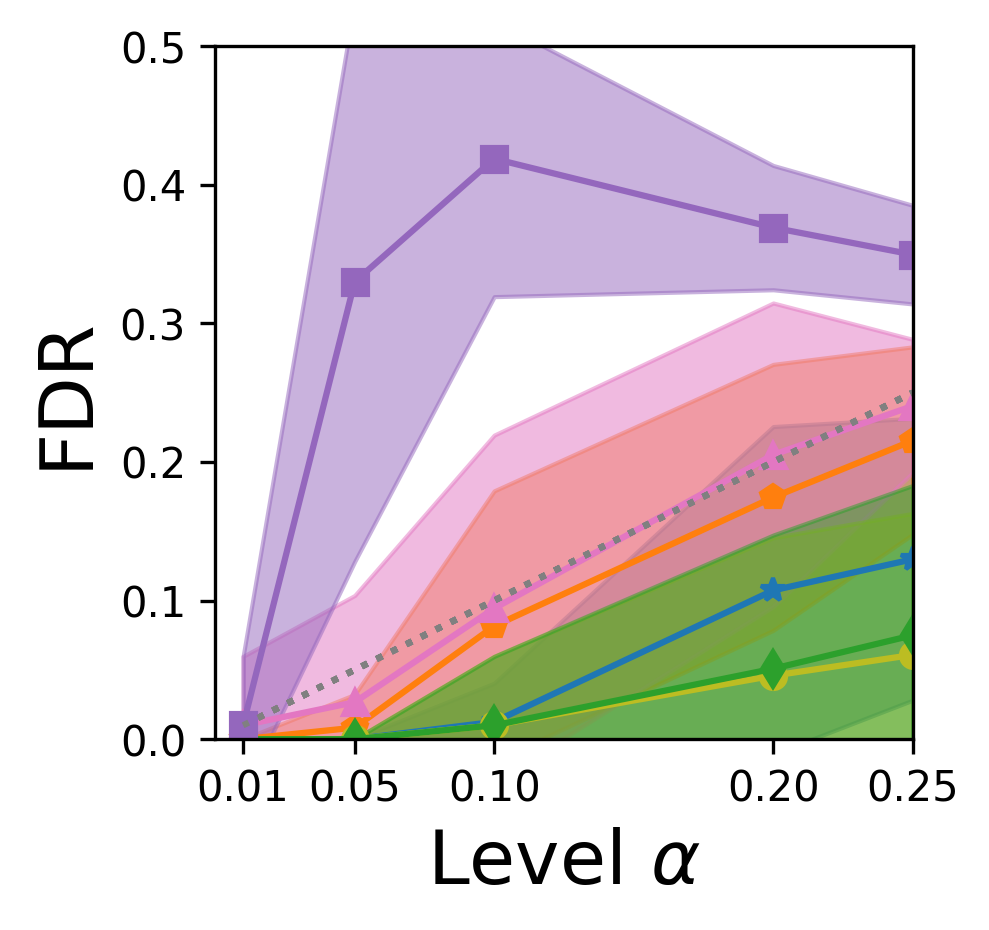}
\includegraphics[width=0.15\linewidth]{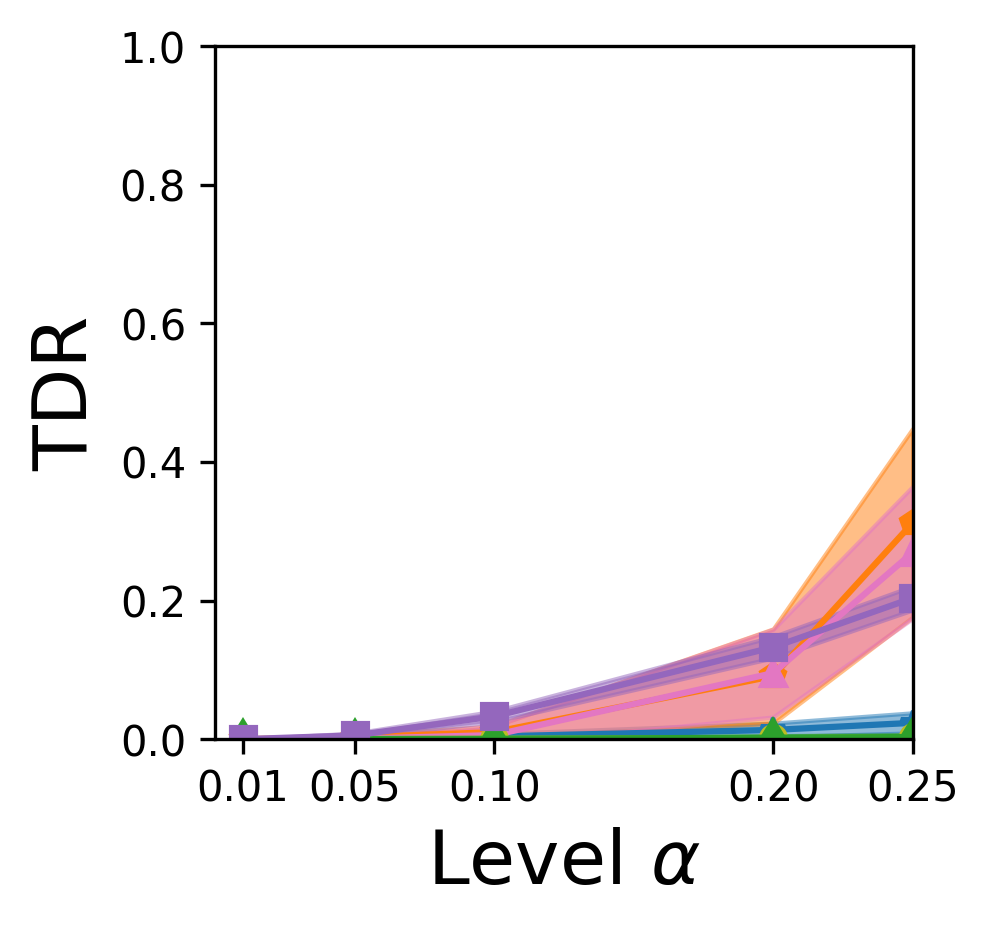}
&
\includegraphics[width=0.15\linewidth]{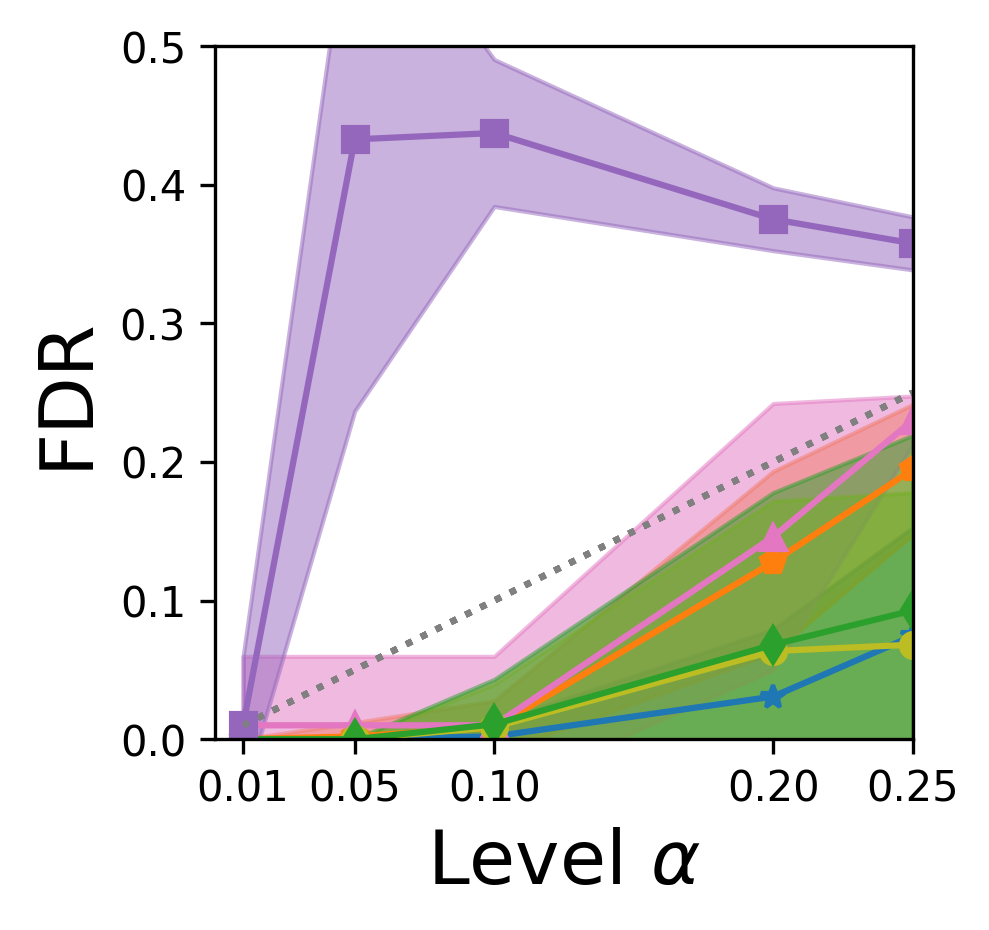}
\includegraphics[width=0.15\linewidth]{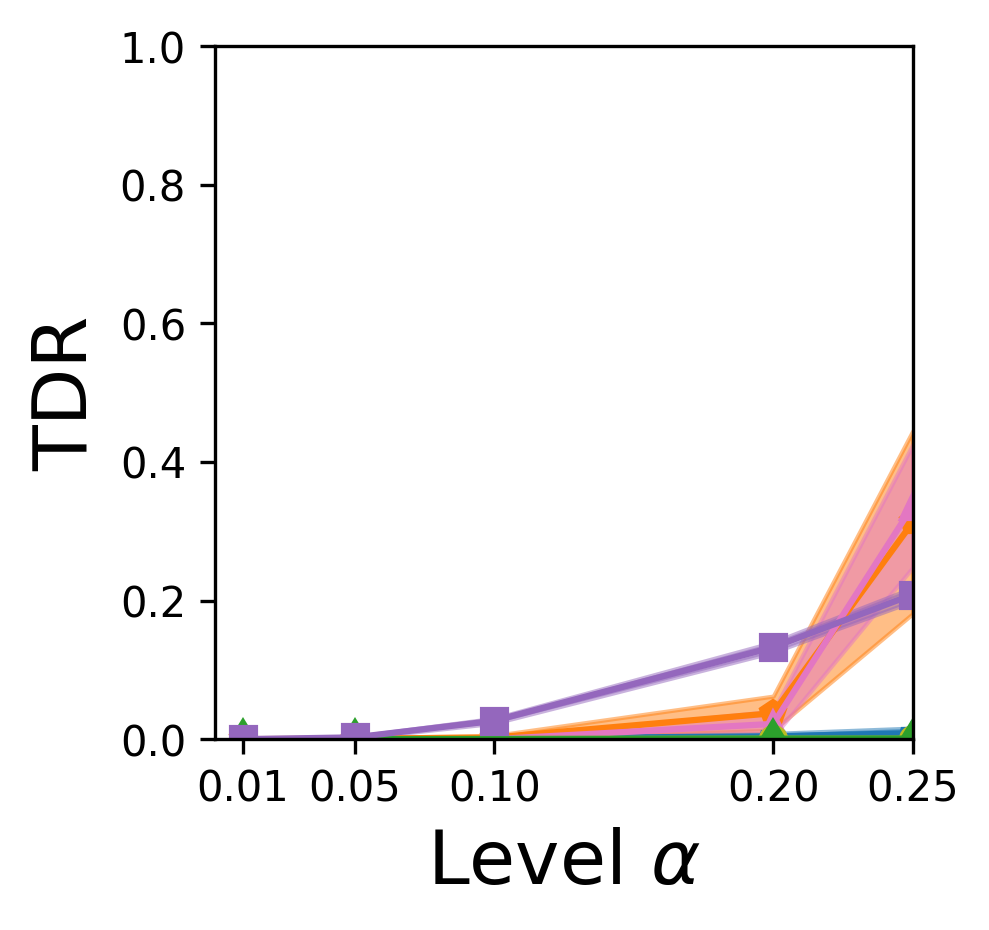}\\
$n=50$ & $n=100$ & $n=200$ \\
\multicolumn{3}{l}{ (2) $p=0.5, p/\epsilon=5$ } \\
 \includegraphics[width=0.15\linewidth]{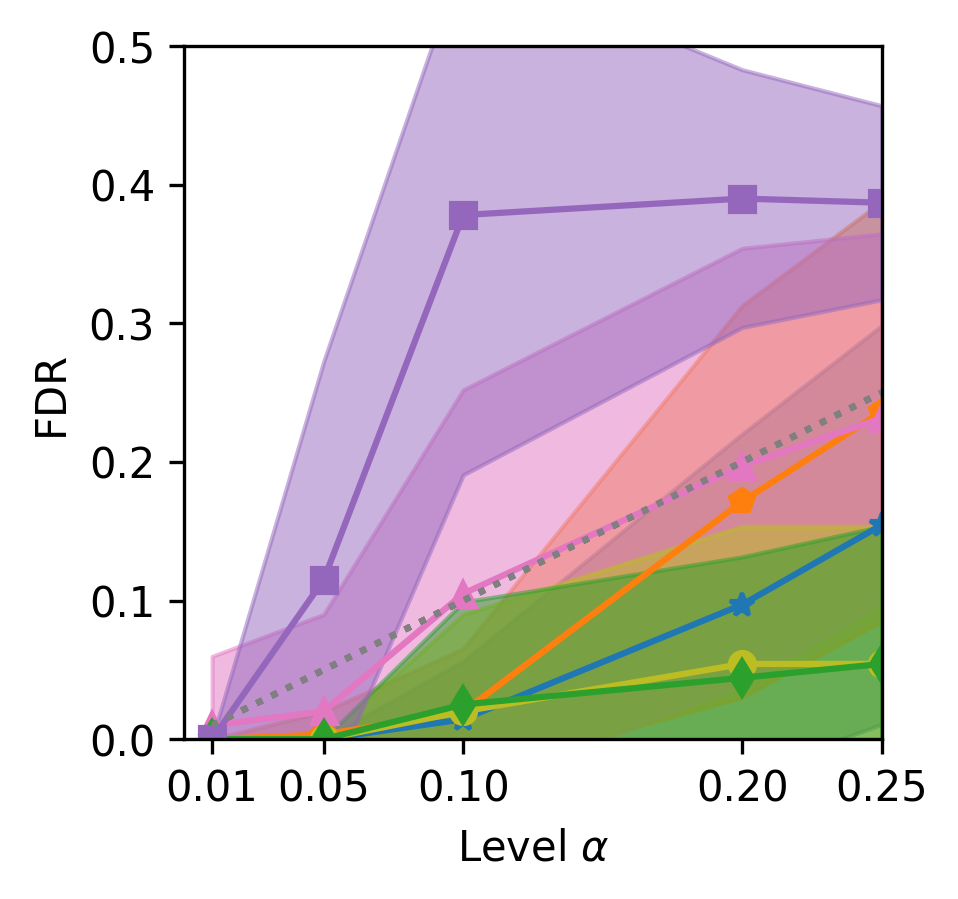}
\includegraphics[width=0.15\linewidth]{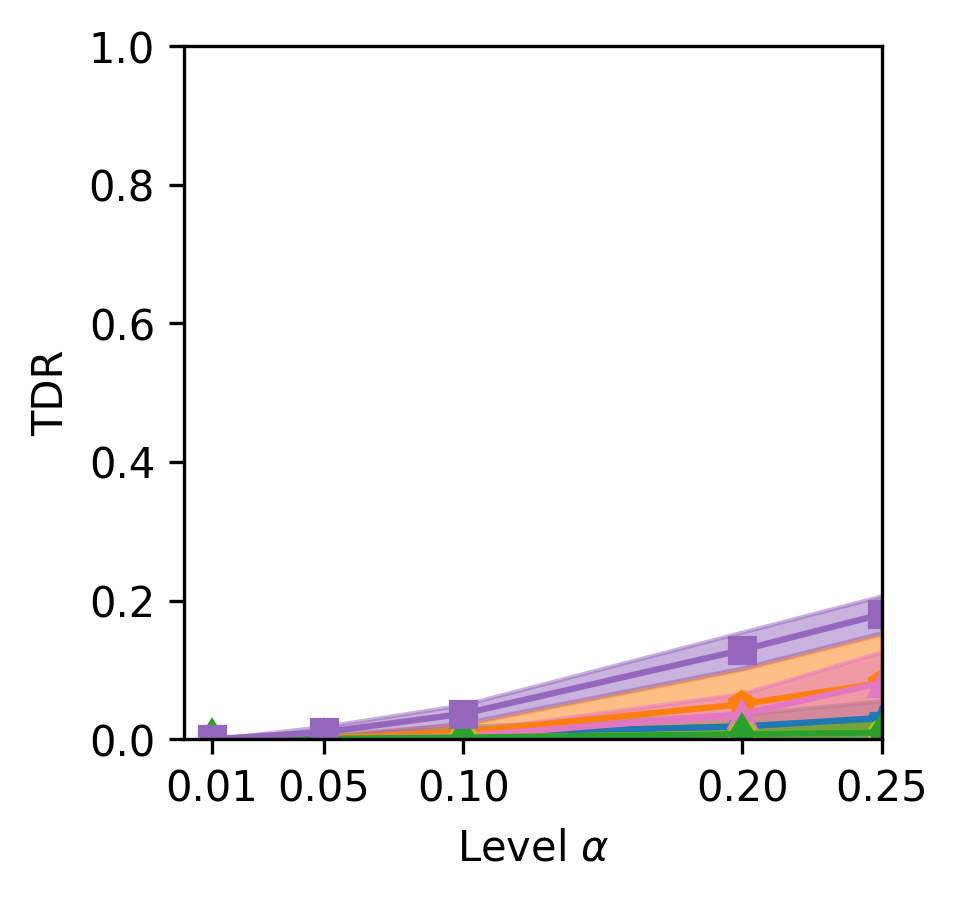}
& 
\includegraphics[width=0.15\linewidth]{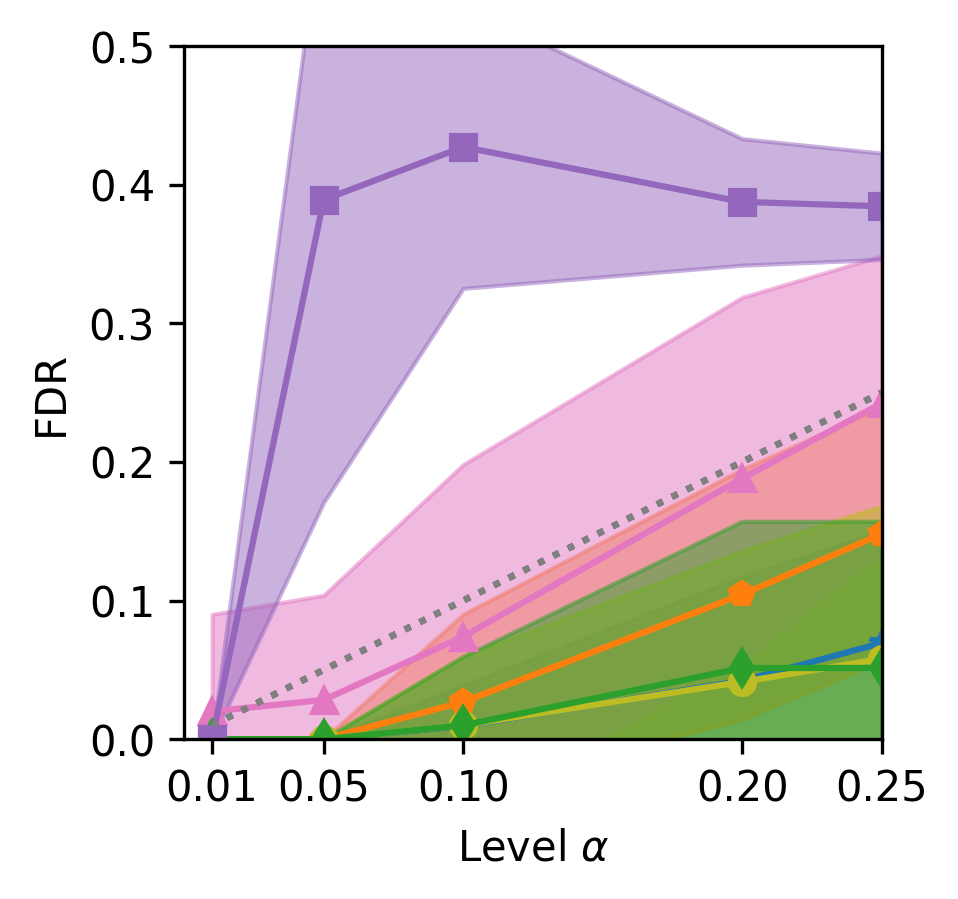}
\includegraphics[width=0.15\linewidth]{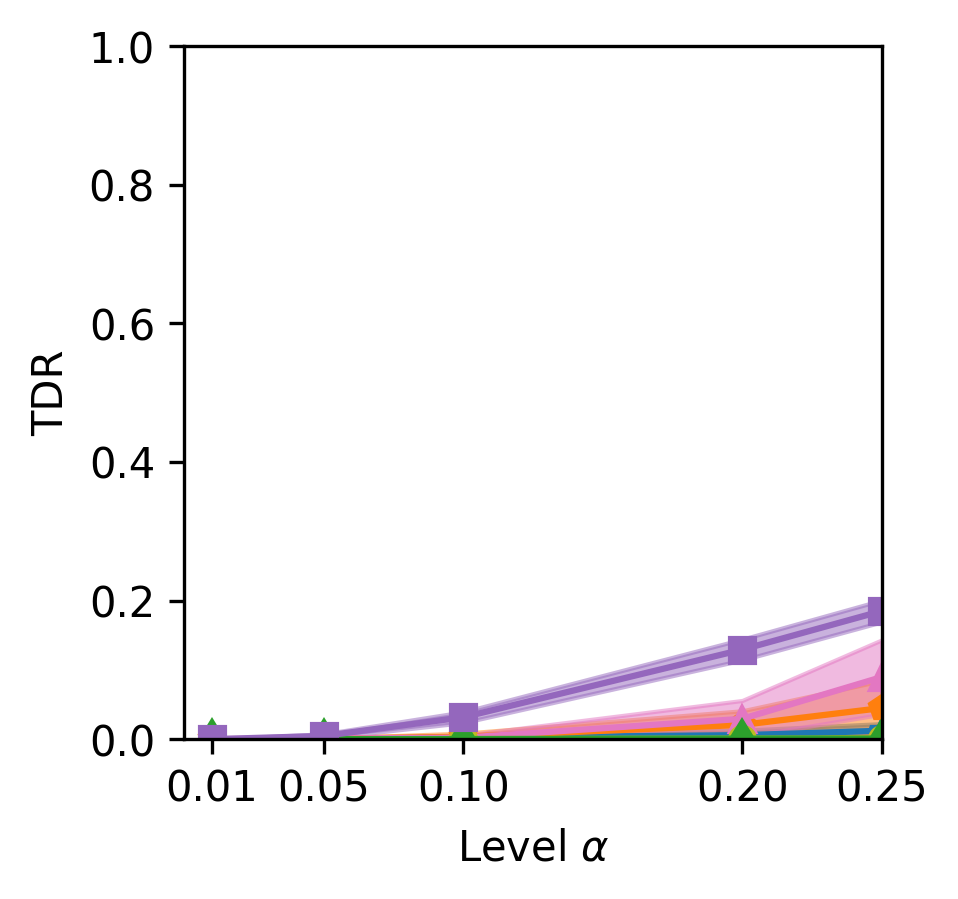} 
&
\includegraphics[width=0.15\linewidth]{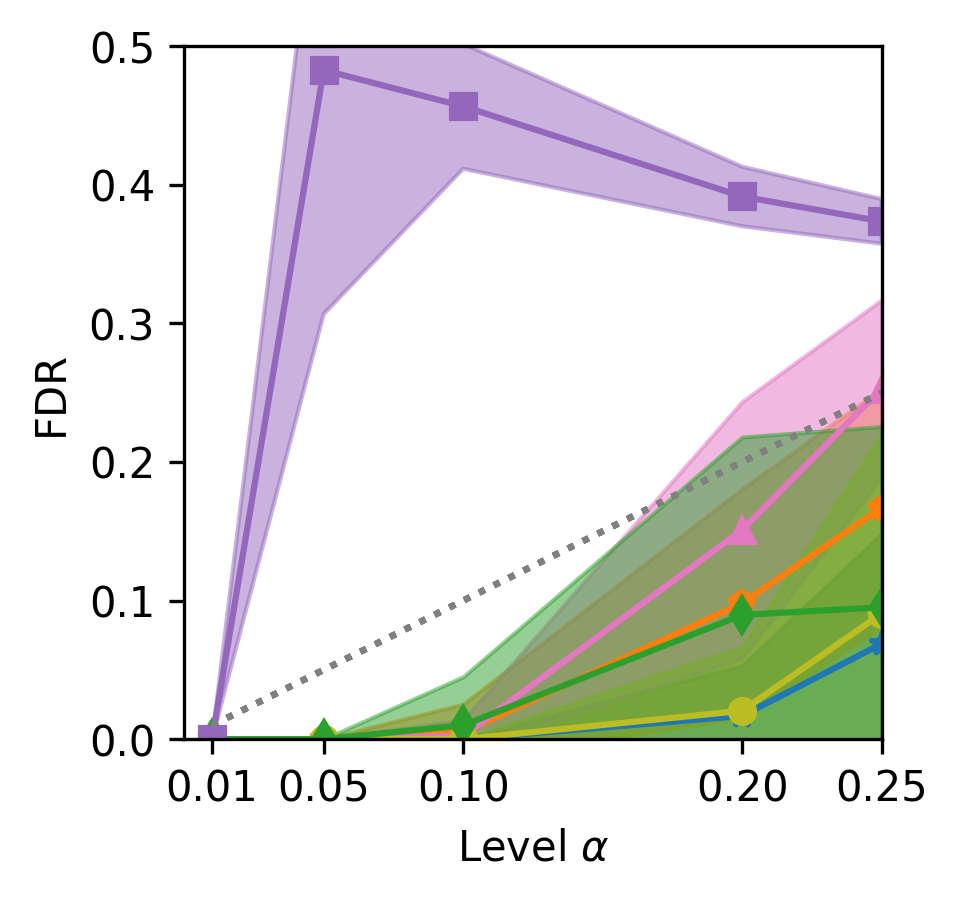}
\includegraphics[width=0.15\linewidth]{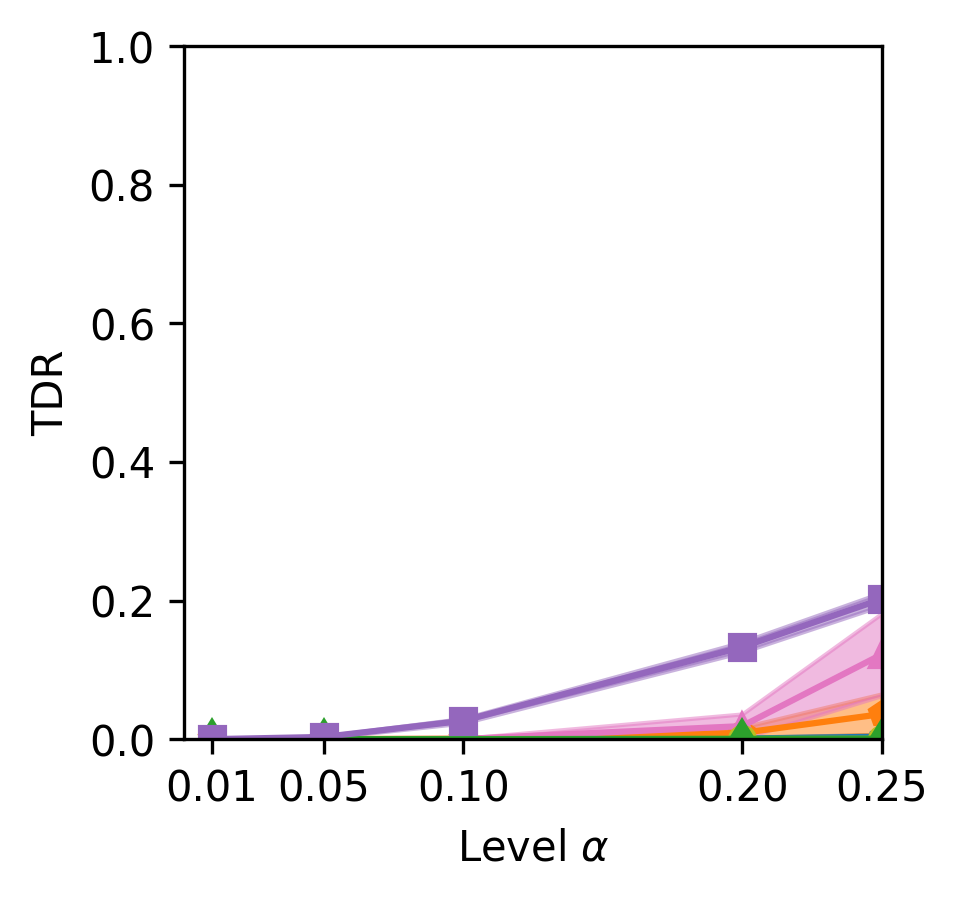} \\

$n=50$ & $n=100$ & $n=200$ \\
\multicolumn{3}{l}{ (3) $p=0.2, p/\epsilon=10$ } \\
 \includegraphics[width=0.15\linewidth]{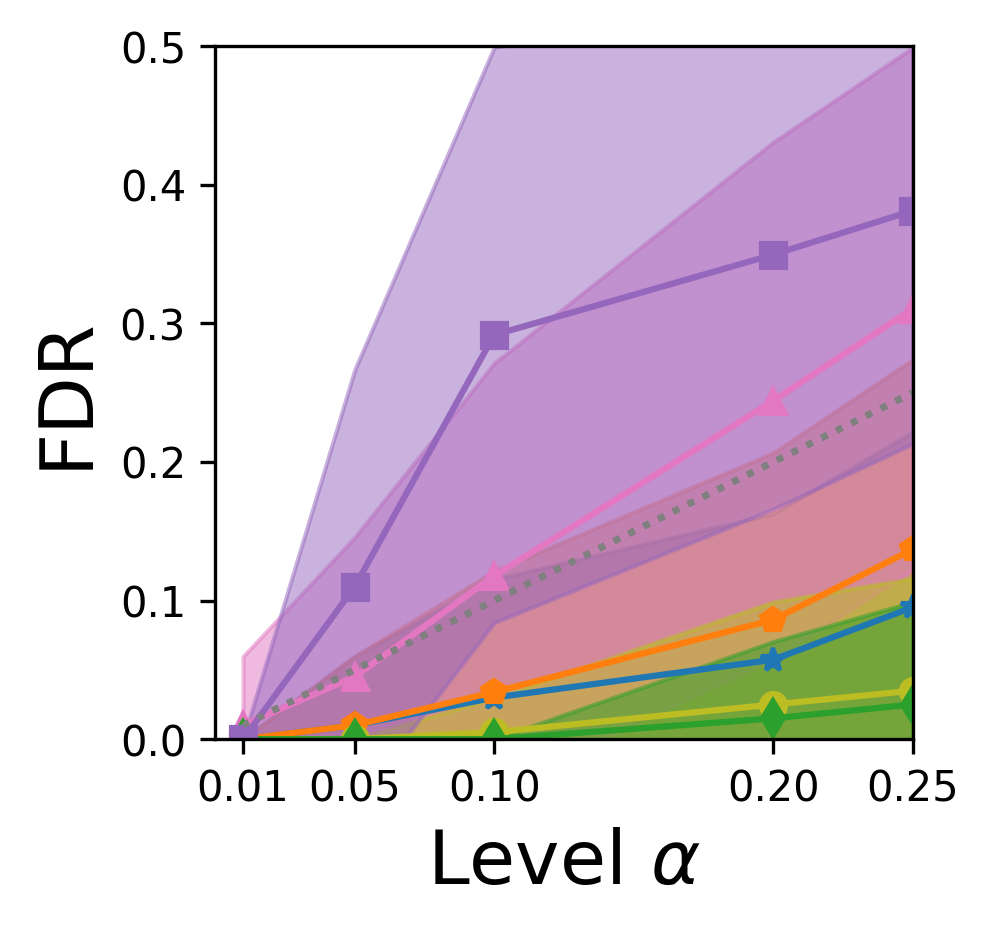}
\includegraphics[width=0.15\linewidth]{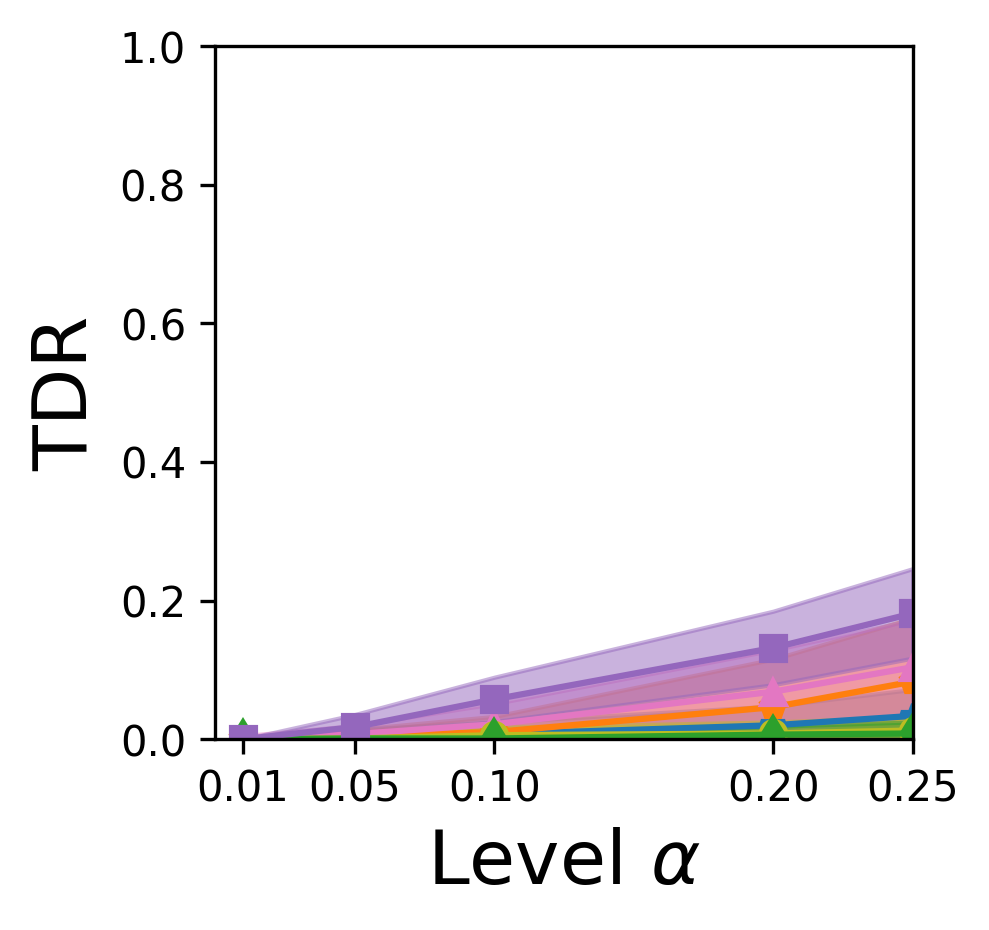}
& 
\includegraphics[width=0.15\linewidth]{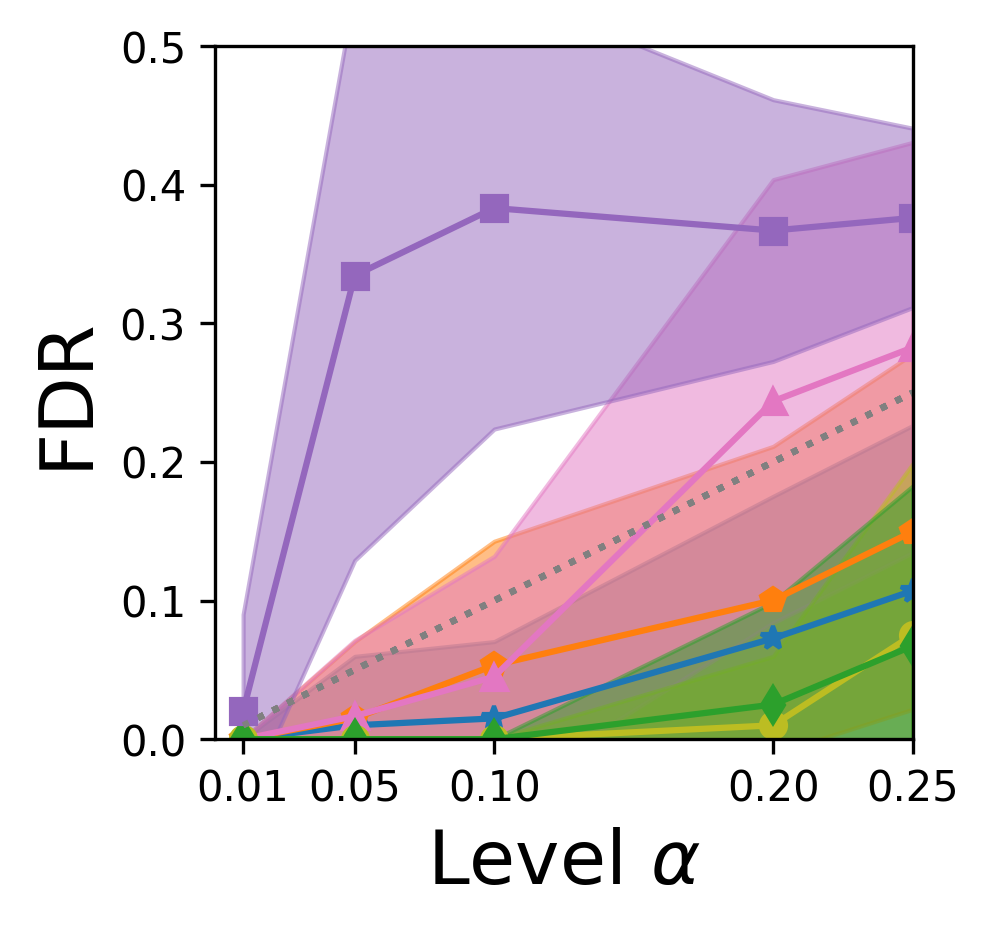}
\includegraphics[width=0.15\linewidth]{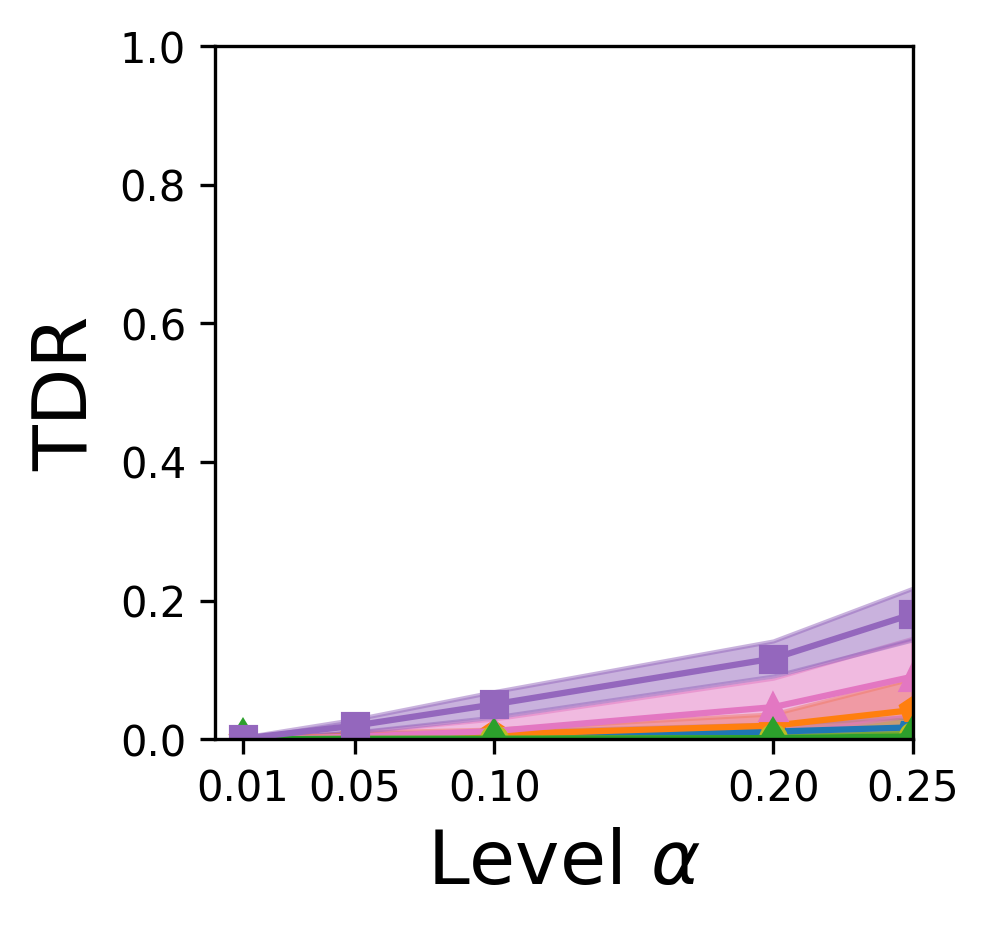} 
&
\includegraphics[width=0.15\linewidth]{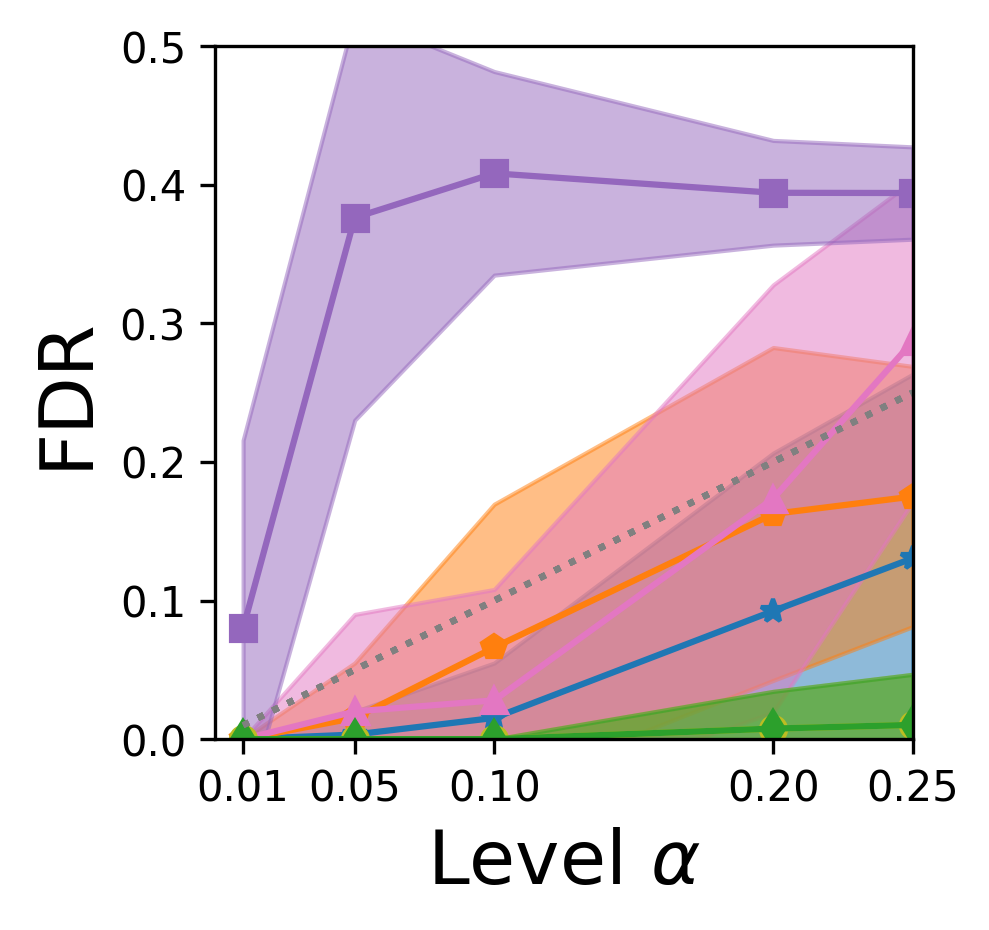}
\includegraphics[width=0.15\linewidth]{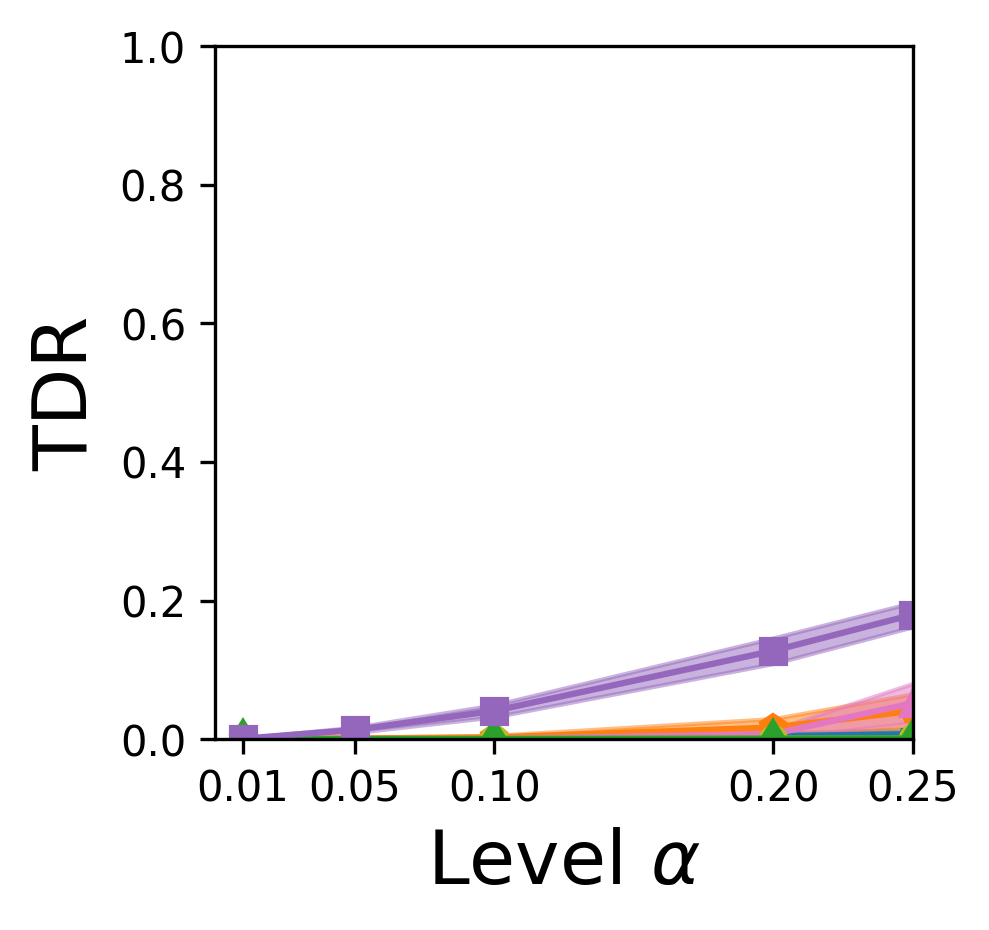} \\

$n=50$ & $n=100$ & $n=200$ \\
\multicolumn{3}{c}{\includegraphics[width=0.8\linewidth]{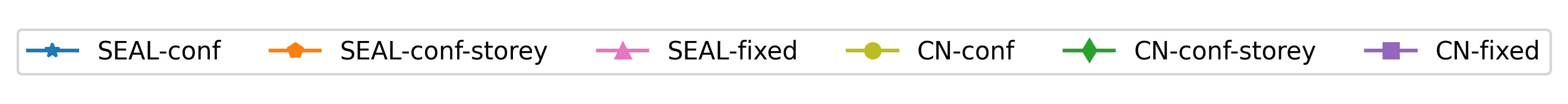}} 
\end{tabular}
\caption{FDR (left panel) and TDR (right panel) as a function of the nominal level $\alpha$. The bands indicate the standard deviation.}
\label{lp:fig:simu:sbm}
\end{figure}

Hence, in order to study the performance of the methods in various conditions, we vary the number of nodes $n$, the SNR $p/ \epsilon$, and the connectivity parameter $p$ (controlling the density $\rho$ for a fixed SNR and a fixed number of nodes $n$). 
In each setting, we construct samples $\cD(Z)$ and test samples $\Dtest(Z)$ 
by removing at random $1-w_1 = 10\%$ of the true edges, and $1-w_0\%$ of the false edges such as to have that $\vert \cH_0 \vert / \vert \cH_1 \vert = 50\%$, and we use $\vert \Dcal \vert = \max( \vert \cD^0 \vert - \vert \cD^1 \vert, 5000)$ in Algorithm \ref{lp:algo:main}. The FDR and TDR of the different methods are evaluated by using 100 Monte-Carlo replications. 
The results are displayed in Figure \ref{lp:fig:simu:sbm} for 1) $p=0.5, p/\epsilon=10$, 2) $p=0.5, p/\epsilon=5$, and 3) $p=0.2, p/\epsilon=10$, with $n \in \{50, 100, 200 \}$.

In this SBM, for any parameter values, the connection probabilities cannot be consistently well estimated by a CN heuristic, as some classes have a low probability of connection within their class while being well connected with other classes: 
for nodes that belong to these groups, it occurs that they share neighbors despite not being connected. Hence, in the case that CN is used, the fixed procedure fails to control the FDR in Figure \ref{lp:fig:simu:sbm}. By contrast, our conformal procedures displays a FDR that is below or close to $\alpha$ across all level values, for all choices of scoring function and for all model configurations. 

However, in some model configurations (when $p =0.2$ or SNR $= 5$) the power is near zero for any scoring function across all values of $\alpha$ and $n$. 
Moreover, even in the more favorable setting where $p=0.5$ and SNR $=10$, the power decreases as $n$ increases if $\alpha$ is low enough ($\alpha \leq 0.2$). 
The issue is that the FDR control problem is intrinsically difficult for a model such as the SBM. To illustrate, consider an oracle setting where the true classes $(C_i)_{1\leq i \leq n}$ and model parameters are known. In that case, a natural strategy is to declare edges using for each test pair $(i,j)$ the probability $\P (A^*_{i,j}=1 \vert (C_i)_{1\leq i \leq n} )$, and if $p > \epsilon$ then the pairs that are most likely to be connected are pairs that are in the same class. However, within a class all pairs have the same likelihood of being connected $\P_\theta (A^*_{i,j}=1 \vert (C_i)_{1\leq i \leq n} ) $ and so if $\alpha < p$, it is impossible to control the FDR at $\alpha$. 
On principle, one can get a finer test statistic than $\P (A^*_{i,j}=1 \vert (C_i)_{1\leq i \leq n} )$ by using the observation $A$, but if the sampling is homogeneous for true and false edges respectively, as is assumed here, $A$ does not give much more information about the location of true and false edges within a class. 
Thus, when $n$ increases, the learned scores concentrate around the oracle probabilities $\P (A^*_{i,j}=1 \vert (C_i)_{1\leq i \leq n} )$, entailing that if $\alpha$ is too low we get near-zero power. When $n$ is small, the probability estimates are more noisy and the margin of error $\alpha$ may be fully utilized by the conformal procedures, which could explain the better power observed here -- however the variance of the FDP naturally increases as $n$ decreases.  
Nonetheless, across all settings, the most powerful procedure among the ones that control the FDR is a conformal method.

\subsubsection{Graphon} \label{lp:sec:simu:graphon}

\begin{figure}
\centering
 \begin{tabular}{ccc}
  \multicolumn{3}{l}{ (1) $\sigma=0.01, a=2$ } \\
 \includegraphics[width=0.15\linewidth]{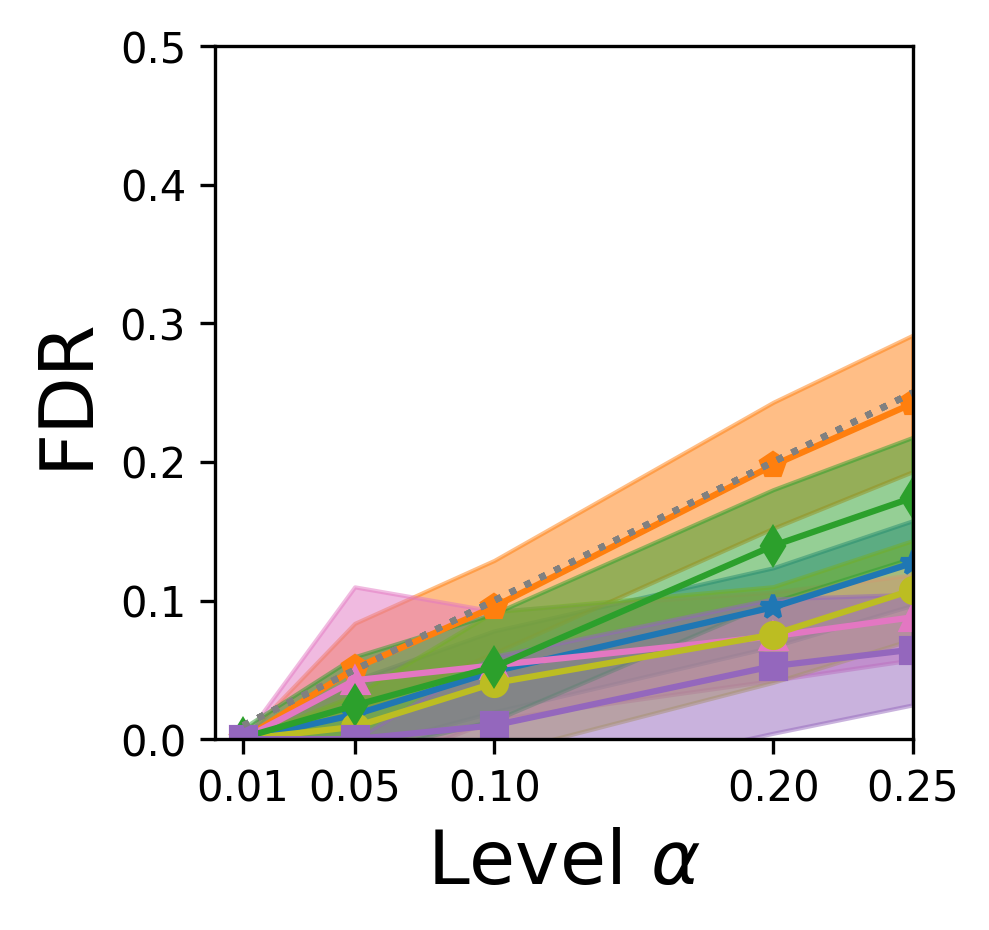}
\includegraphics[width=0.15\linewidth]{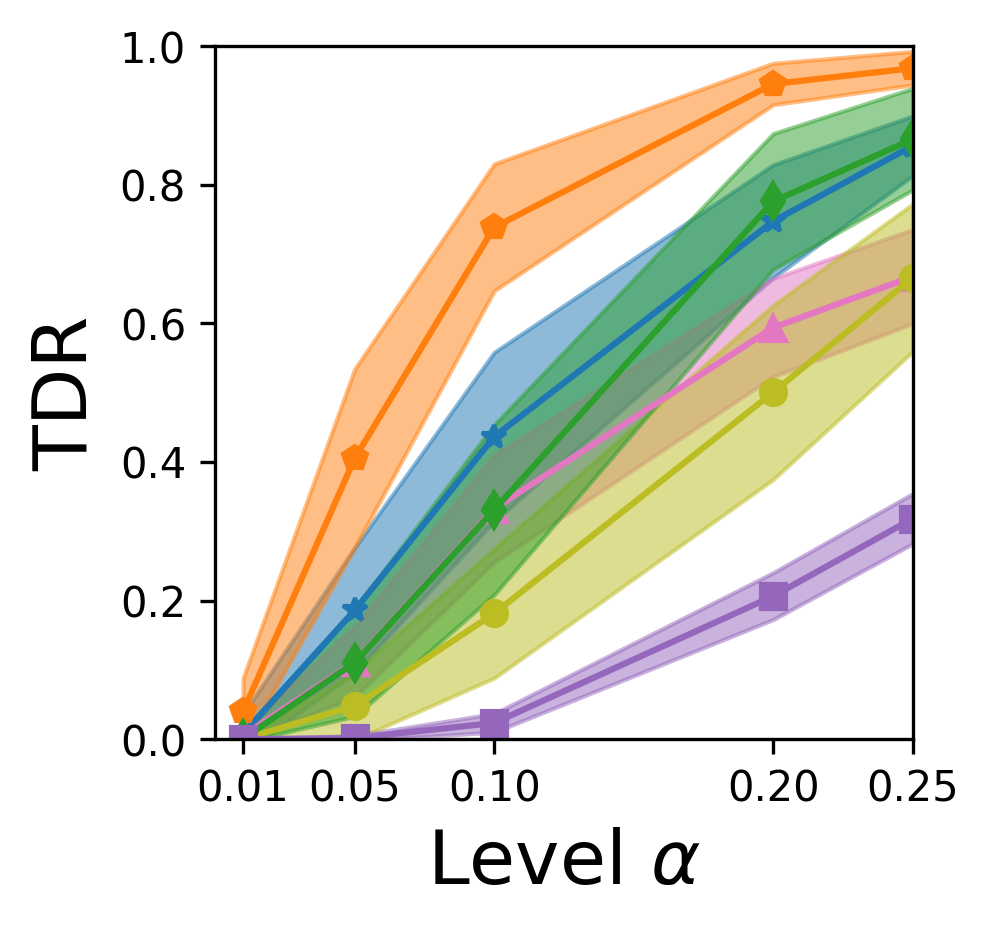}
&
\includegraphics[width=0.15\linewidth]{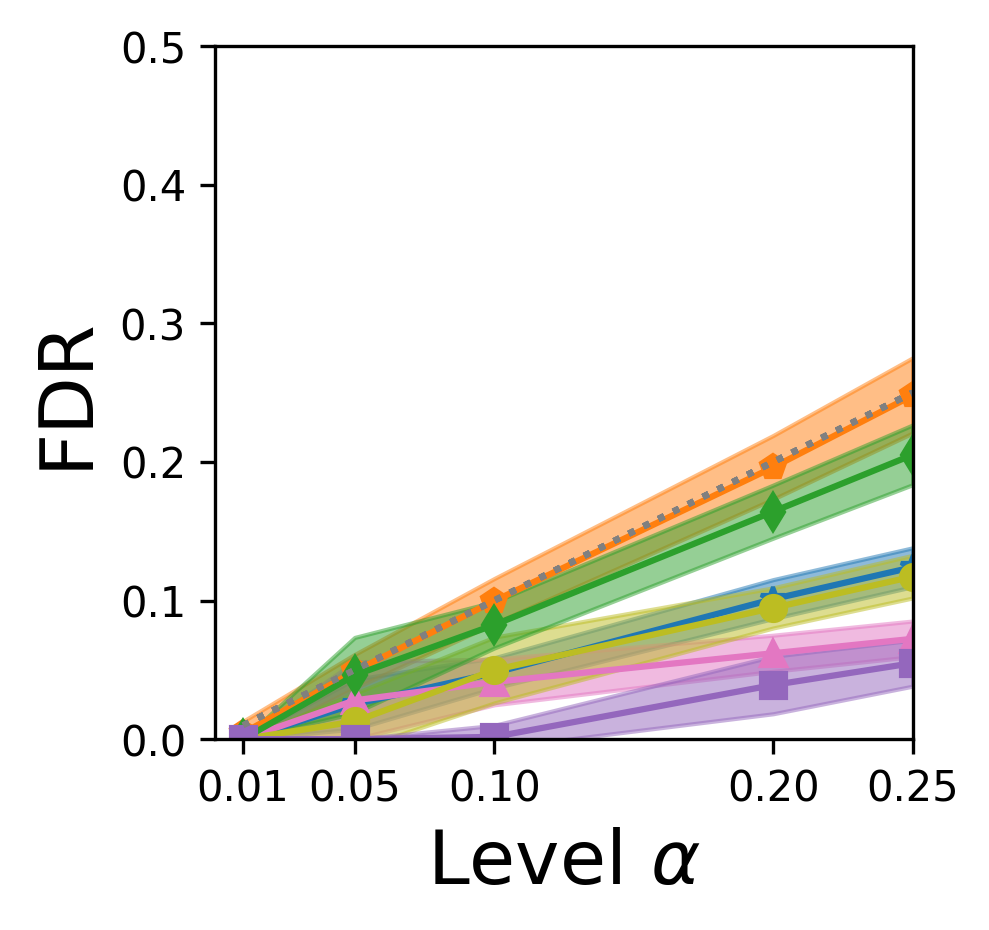}
\includegraphics[width=0.15\linewidth]{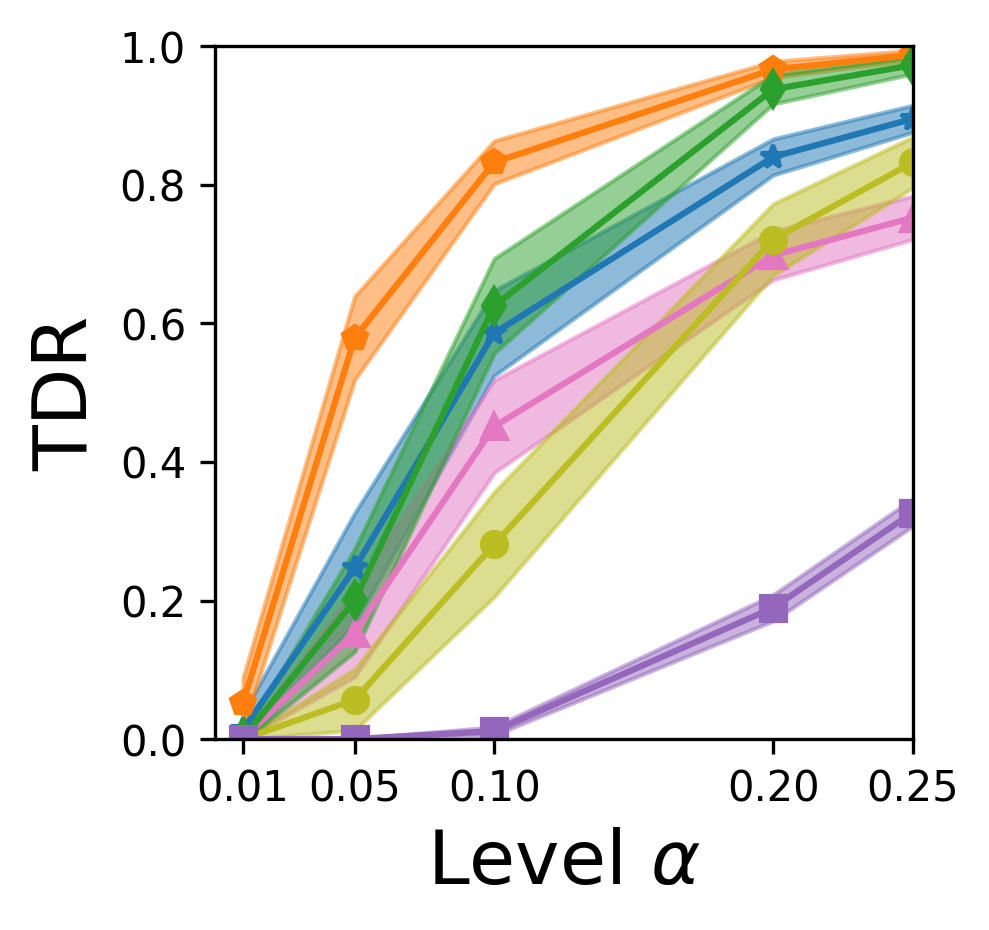}
&
\includegraphics[width=0.15\linewidth]{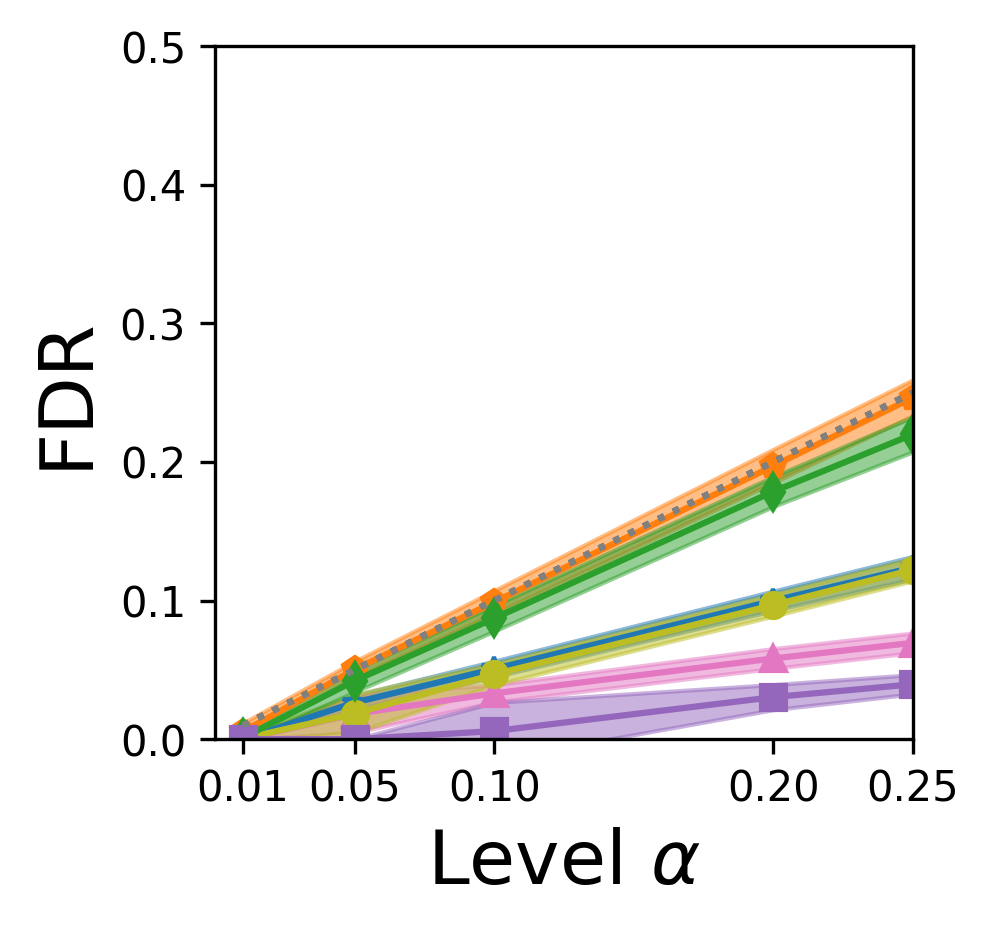}
\includegraphics[width=0.15\linewidth]{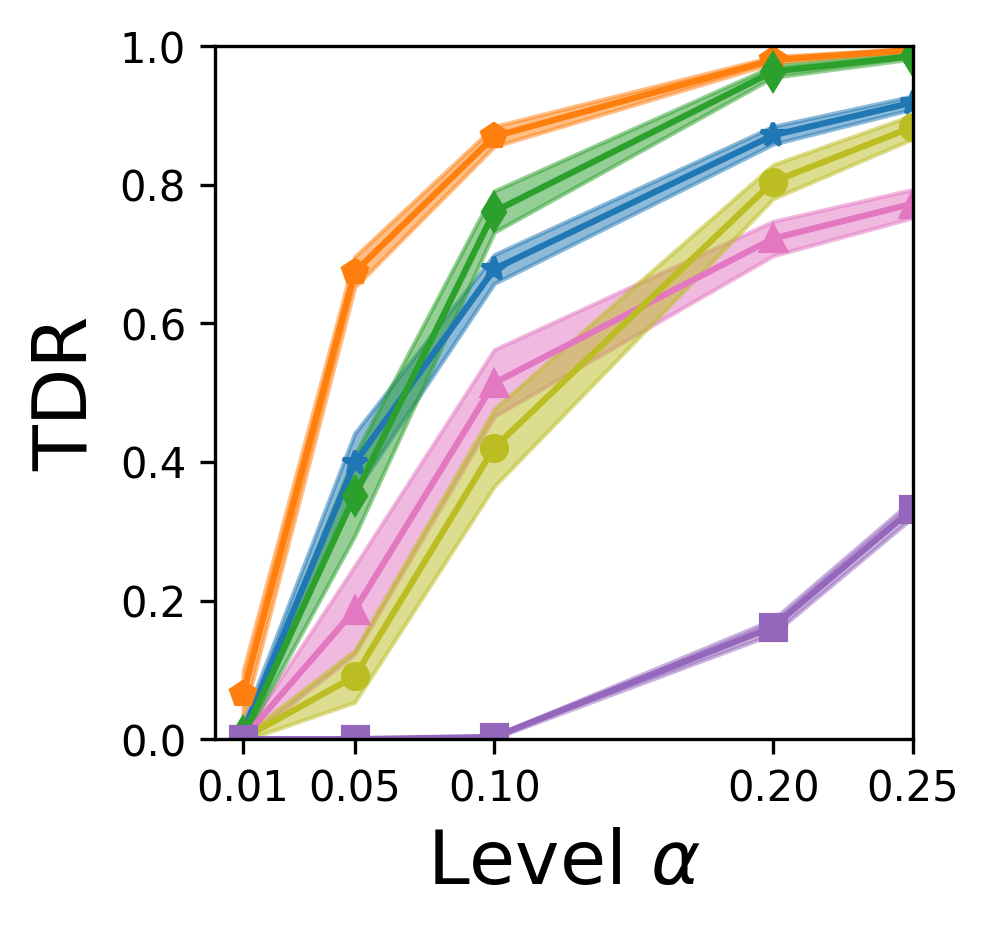}\\
$n=50$ & $n=100$ & $n=200$ \\
 \multicolumn{3}{l}{ (2) $\sigma=0.1, a=0.5$ } \\
 \includegraphics[width=0.15\linewidth]{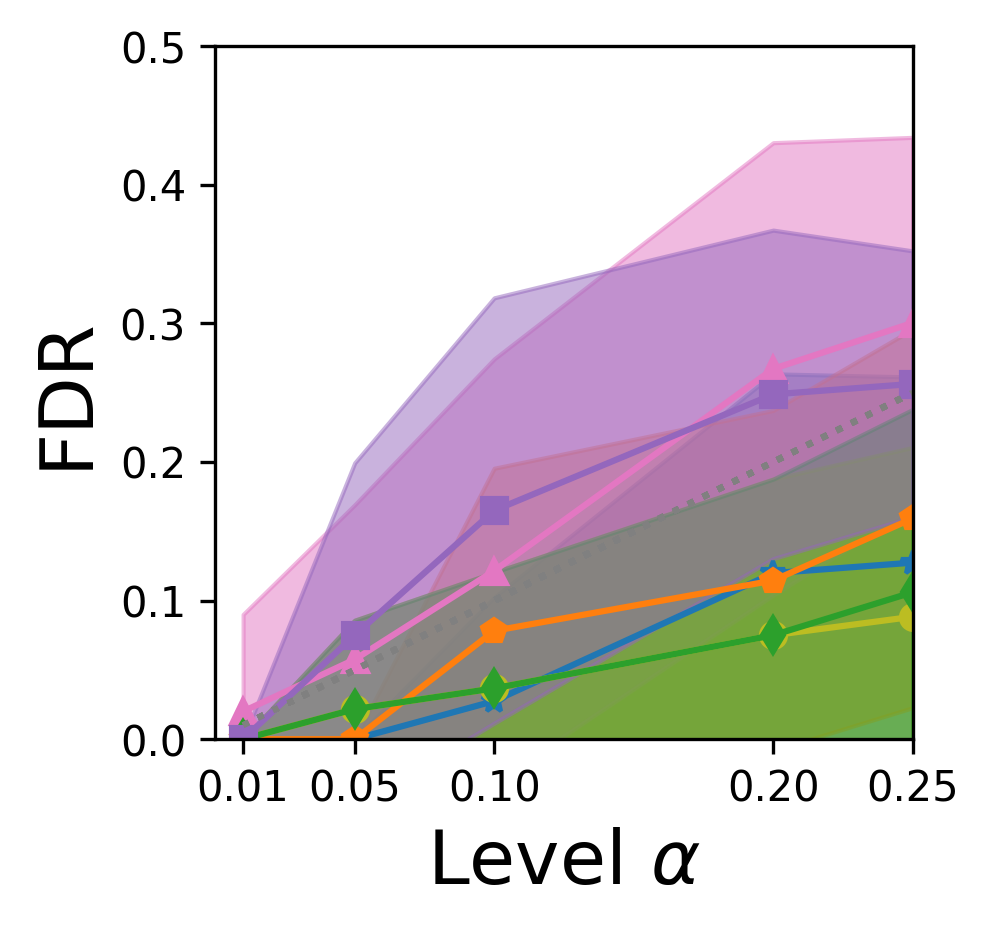}
\includegraphics[width=0.15\linewidth]{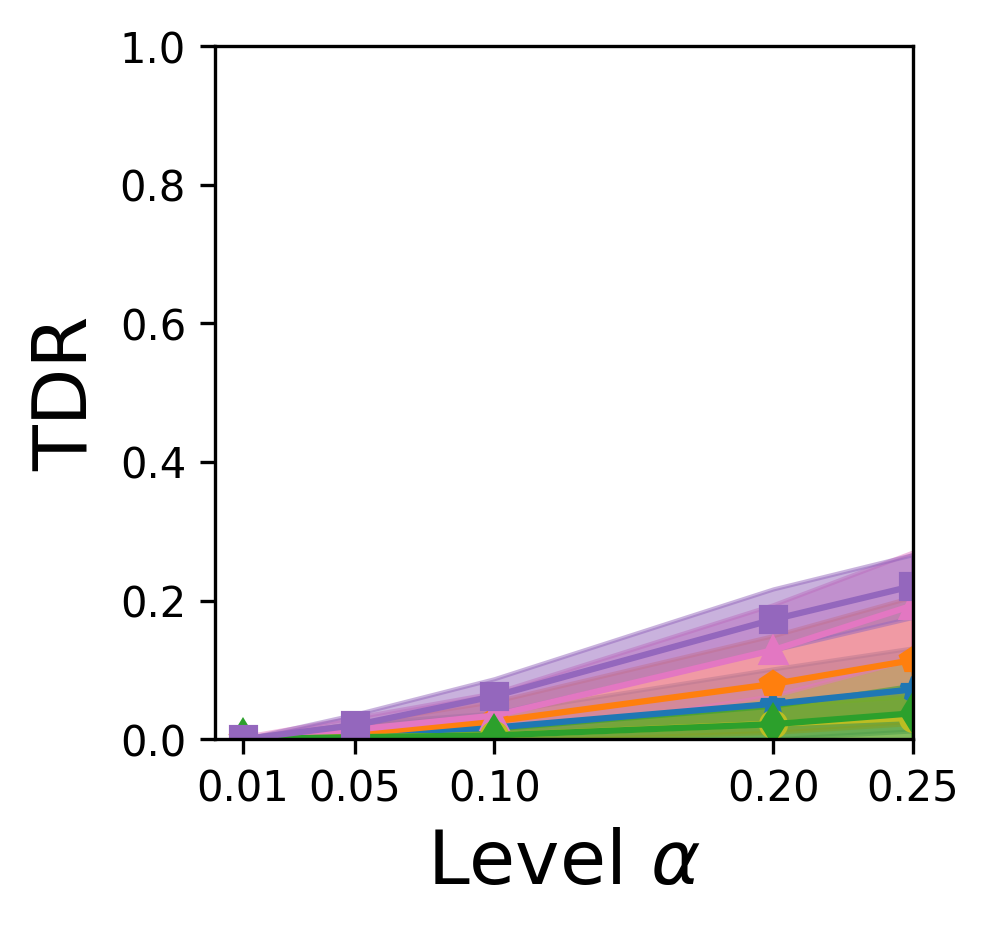}
&
\includegraphics[width=0.15\linewidth]{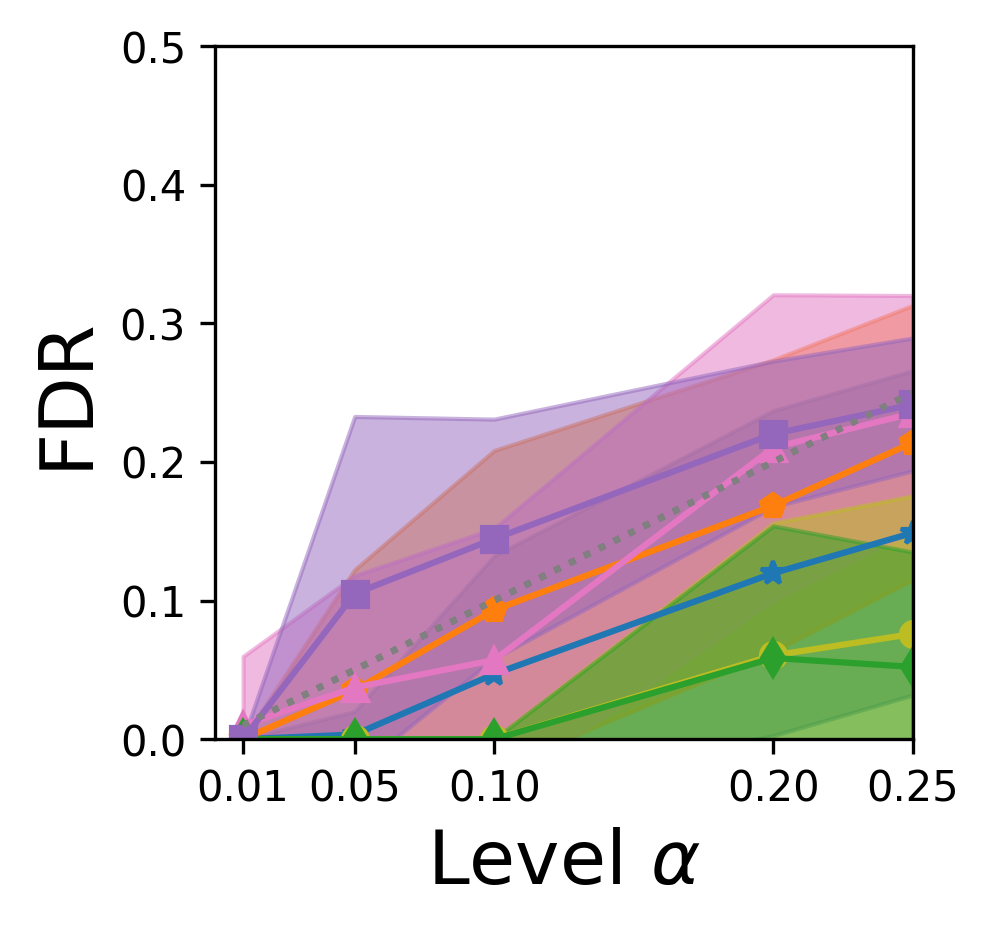}
\includegraphics[width=0.15\linewidth]{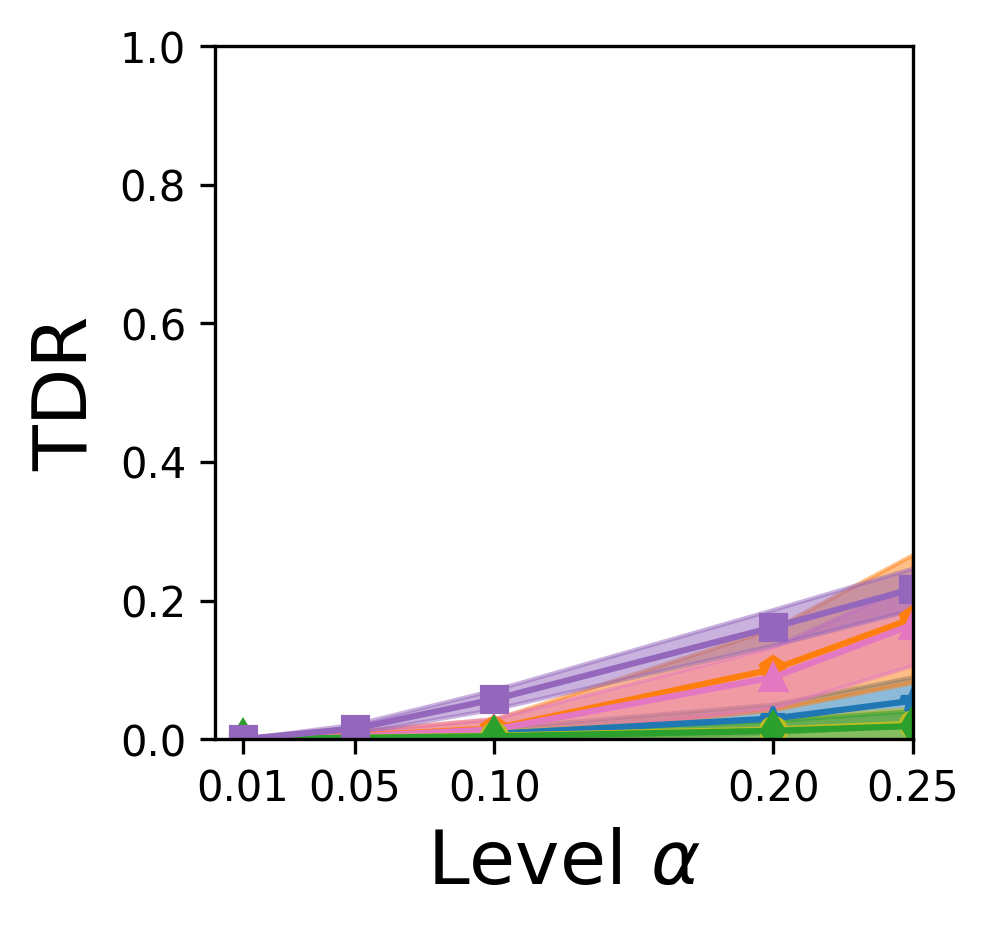}
&
\includegraphics[width=0.15\linewidth]{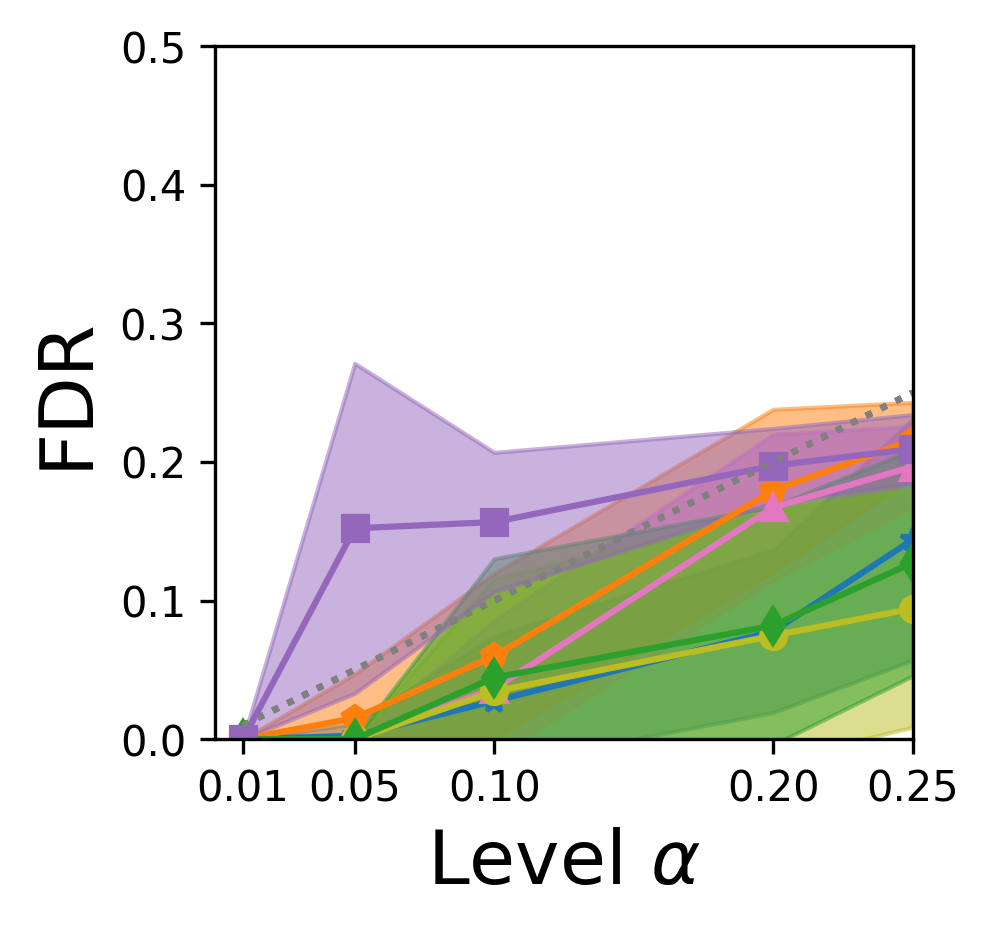}
\includegraphics[width=0.15\linewidth]{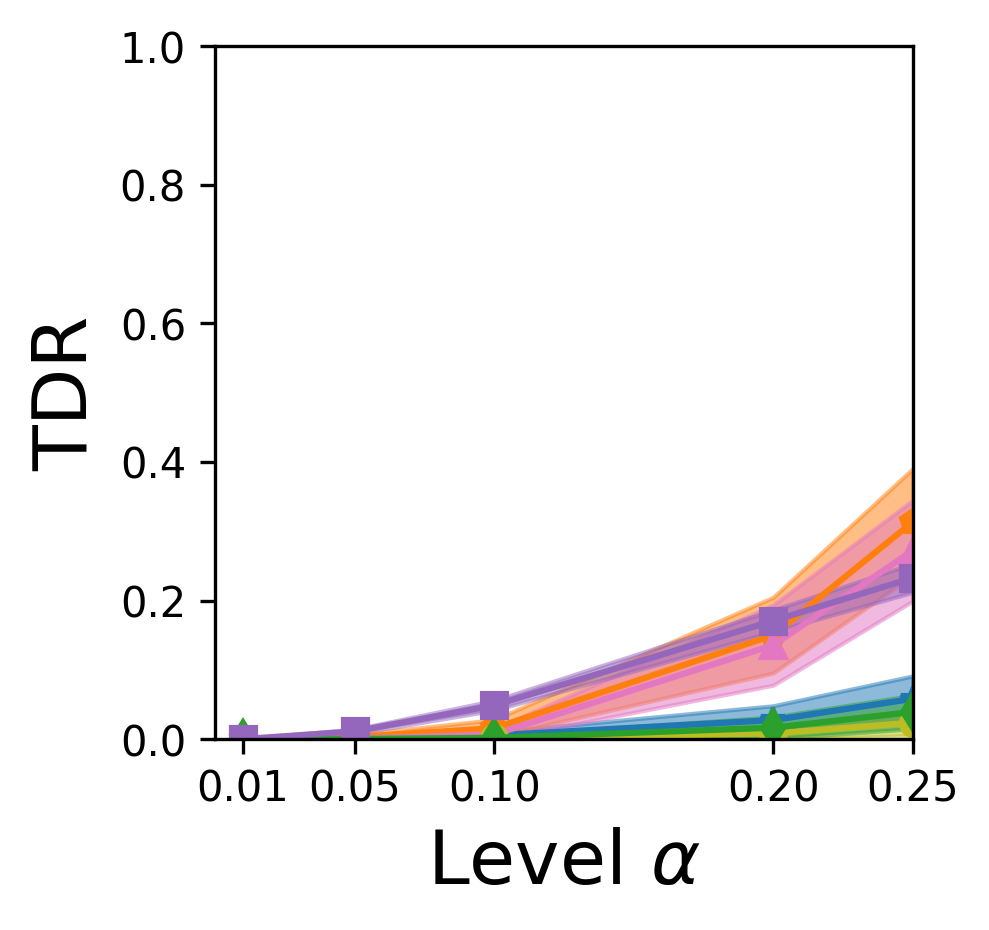} \\
$n=50$ & $n=100$ & $n=200$ \\
\multicolumn{3}{c}{\includegraphics[width=0.8\linewidth]{legend.png}} 
\end{tabular}
\caption{FDR (left panel) and TDR (right panel) as a function of the nominal level $\alpha$. The bands indicate the standard deviation.}
\label{lp:fig:simu:graphon}
\end{figure}

Next, $A^*$ is generated using a graphon model:
\begin{align*}
\xi_1, \dots, \xi_n &\underset{i.i.d.}{\sim} U[0,1] \\
A^*_{i,j} \vert \xi_1,  \dots, \xi_n &\sim \cB \left( \text{exp} (-(2 \sigma)^{-1}(\xi_i - \xi_j)^a ) \right), \quad 1 \leq i < j \leq n . 
\end{align*}
We construct samples $\cD(Z)$ and test samples $\Dtest(Z)$ and choose $\vert \Dcal \vert$ as in the previous section. The FDR and TDR of the methods are displayed in Figure \ref{lp:fig:simu:graphon} for $\sigma=0.01, a=2$ and for $\sigma=0.1, a=0.5$, with $n \in \{50, 100, 200 \}$.
In this model, compared to Section \ref{lp:sec:simu:sbm}, the CN heuristic is not necessarily a poor predictor: by transitivity, the graphon function implies that the more neighbors two nodes have in common, the more likely it is that they are connected.

In the case that $\sigma=0.01$ and $a=2$, all procedures control the FDR for all $n$ and $\alpha$ values. However, our conformal procedures are the most powerful by a margin. In the more difficult setup where $\sigma=0.1$ and $a=0.5$, the FDR of the fixed procedure is inflated: across all values of $n$ in the case of CN and for small $n$ ($n=50$) in the case of SEAL. By contrast, our conformal procedures control the FDR across all values of $n$ in this setup also, with equal or higher power than the fixed procedure when its FDR is below the nominal level. Finally, for all procedures the power improves when $n$ increases, however the gain is more substantial for the conformal methods.

\subsection{Real data} \label{lp:sec:realdata}

We evaluate our method on a real dataset: a food web network \citep{foodweb99} downloaded from the Web of Life Repository (\url{https://www.web-of-life.es/}). 
A food web is a network of feeding interactions in an ecological community, in which nodes are species and two species are connected if one eats the other. Food web analysis delivers important insights into the workings of an ecosystem. 
However, documenting species interaction is in practice labour-intensive and the resources for such investigations finite; hence, food webs are typically incomplete.

\begin{figure}
\centering
\includegraphics[width=0.5\linewidth]{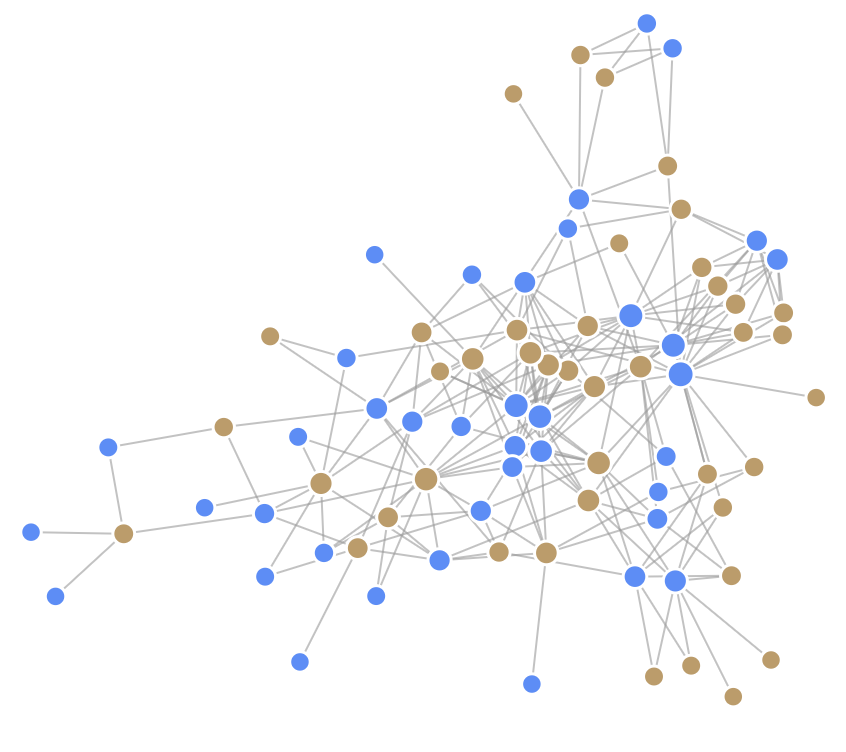}
\caption{Food web network of \cite{foodweb99} with a bipartite representation: prey nodes are colored in blue and predator nodes are colored in yellow.}
\label{lp:fig:fw}
\end{figure}

In this dataset, there are in total 48 species and 221 interactions recorded between them. 
We consider a directed and bipartite representation where a set of predators connect to a set of preys: if a species plays both roles (i.e. eats certain species and also gets eaten by others), then we associate with it two separate nodes, one in the predator set and one in the prey set. We thus obtain a network of 81 nodes and 221 edges (see Figure \ref{lp:fig:fw}). In the corresponding adjacency matrix, there is by construction a zero entry for any node pair $(i,j)$ where $i$ is a prey or both $i$ and $j$ are predators: hence, we remove these node pairs from the set of negative edges $\{(i,j) \colon A^*_{i,j} =0 \}$ in the sequel.

\begin{figure}
\centering
 \includegraphics[width=0.25\linewidth]{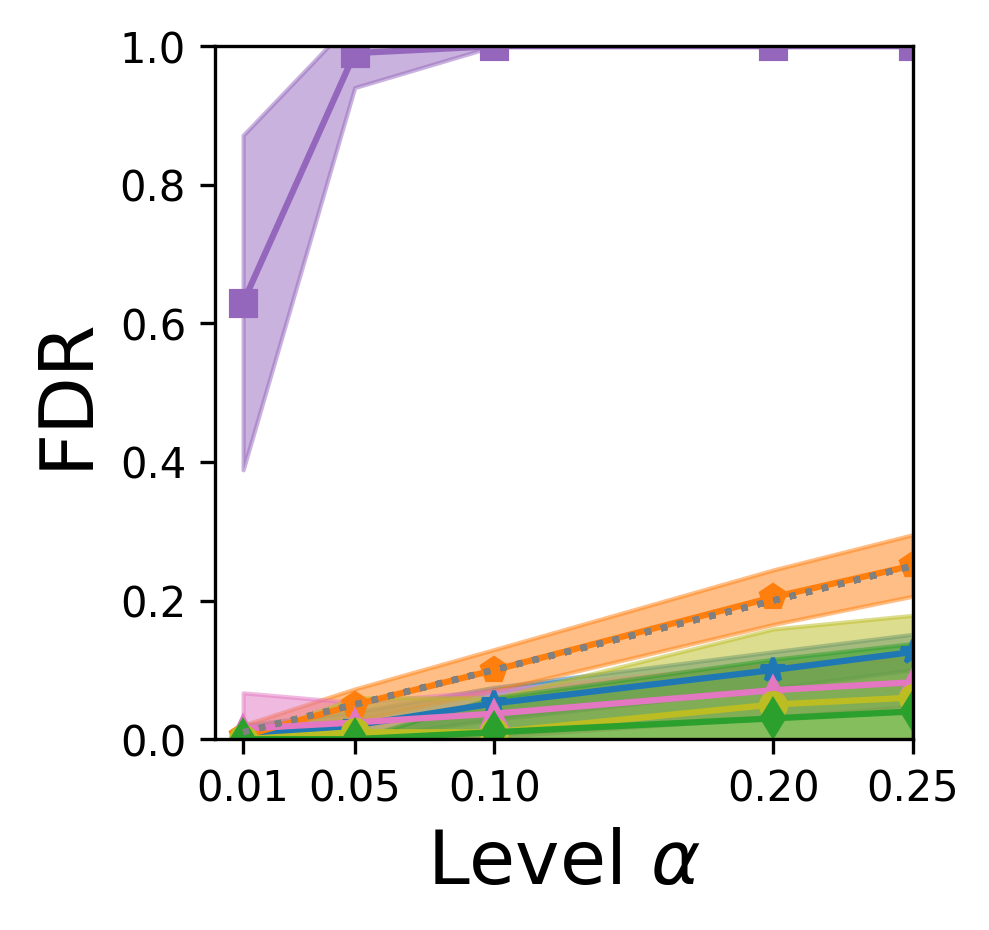}
\includegraphics[width=0.25\linewidth]{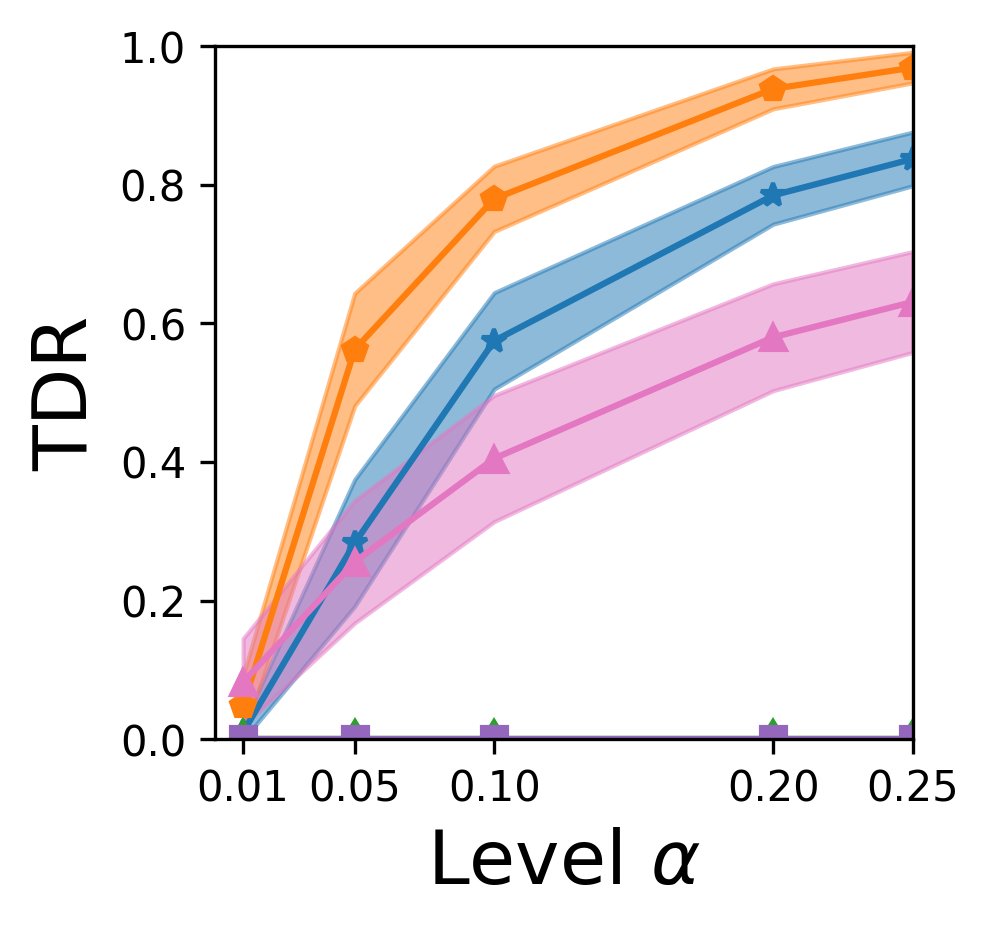} \\
\includegraphics[width=0.8\linewidth]{legend.png}
\caption{FDR (left panel) and TDR (right panel) as a function of the nominal level $\alpha$, for the food web dataset described in Section \ref{lp:sec:realdata}. The bands indicate the standard deviation.}
\label{lp:fig:app}
\end{figure}

We construct samples $\cD(Z)$ and test samples $\Dtest(Z)$ by considering the dataset at hand as the complete network $A^*$ and generating the missingness ourselves, 
using double standard sampling with $w_1=10\%$ and $w_0$ chosen such that $\vert \cH_0 \vert / \vert \cH_1 \vert = 50\%$, and use $\vert \Dcal \vert = \max( \vert \cD^0 \vert - \vert \cD^1 \vert, 5000)$ for the conformal methods. The performance of each method is then evaluated by computing the FDP and TDP with respect to the ground truth and the FDR and TDR are computed by using 100 Monte-Carlo replications.  
The FDR and TDR of the methods are displayed in Figure \ref{lp:fig:app} for varying $\alpha$. 

When the scoring function uses the CN heuristic, the fixed threshold procedure fails to control the FDR. 
Indeed, the bipartite structure entails poor estimation of edge probabilities with such a link prediction heuristic, as prey nodes and predator nodes have, respectively, many neighbors in common but share no connections with each other. Conversely, SEAL is not only learning-based but is also expressive enough that it can learn various connection patterns. Hence, we observe here that when using SEAL, the fixed procedure controls the FDR. 
By contrast, our conformal procedures control the FDR regardless of the link prediction algorithm used for the scoring function and so even in case of a poor link prediction model. 
Moreover, the conformal procedures uniformly improve upon the fixed procedure:  when the fixed threshold procedure has an FDR under the nominal level, our conformal procedures have a much higher power.

\section{Discussion} \label{lp:sec:disc}

We have proposed a novel method that calibrates the output of any link prediction technique for FDR control, using recent ideas from the conformal inference literature. 
In a nutshell, our method acts as a wrapper that allows to transform an off-the-shelf link prediction technique into a procedure that controls the proportion of falsely detected edges at a user-defined level.
Importantly, our proposed method is model-free: no assumptions are made regarding the distribution of the complete graph and the control is provided regardless of the quality of the link prediction model. 

In this work, it is assumed that the sampling matrix is generated by sampling at random true and false edges at an unknown rate respectively. However this type of sampling does not include certain practical sampling designs such as egocentric sampling \citep{levina23}. On the other hand, the key of the method is to sample the calibration node pairs in a manner that mimics the missingness of the false edges in the test set: therefore, if the sampling mechanism is complicated but its parameters are known, the control can still provided as long as the sampling of the calibration sample is adapted.

An important topic in conformal inference is the question of conditional guarantees, that is, achieving error guarantees that hold conditionally on somes component of the data. For instance, in classical conformal prediction, a large body of work \citep{foygel2021limits, romano2020classification, romano2019conformalized} studies the aim of obtaining valid prediction sets conditional on the features of a new test point, while a few others \citep{lofstrom2015bias, sadinle2019least} are rather interested in conditioning on the label. In the outlier detection context, \cite{bates2023testing} considers guarantees conditional on the calibration data (note that labels are also fixed in that work). In our case, let us note that the control can be considered as conditional on the complete graph $A^*$: indeed, the calibration node pairs are actually sampled in a manner that mimics the missingness of the false edges conditionally on $A^*$. Hence, the randomness of the problem resides in the sampling matrix. However, defining a relevant variable to further condition on is not clear here, since the observed data has a complex dependence structure.

Finally, some related scenarios of interest are settings where some true/false edges are erroneously observed or where the sampling matrix is not observed \citep{levina17}.
Investigating the extension of the method for these cases represents an interesting avenue of research for future work.  

\section*{Acknowledgements}
The authors would like to thank Tabea Rebafka, Etienne Roquain, Nataliya Sokolovska, Gilles Blanchard, Guillermo Durand and Romain Perier for their constructive feedbacks. A. Marandon has been supported by a grant from R\'egion \^Ile-de-France (``DIM Math Innov'').

\bibliographystyle{apalike}
\bibliography{bibli_lp}

\end{document}